# German Cities with Universities: Socioeconomic Position and University Performance

Anthony F.J. van Raan

Centre for Science and Technology Studies, Leiden University, Leiden, The Netherlands
vanraan@cwts.leidenuniv.nl
ORCID: 0000-0001-8980-5937

## Abstract

*A much debated topic is the role of universities in the prosperity of cities and regions. When researching this, one is faced with two major problems. First, what is a reliable measurement of the diverse elements of prosperity; and second, given the wide variety of types of universities, what are the characteristics, particularly research performance of a university that really matter? In order to take a step forward in the discussion, we focus in this study on the research question: Is there a significant relation between having a university and a city's socioeconomic strength, growth of gross urban product, and growth of population size? And if so, what are the determining indicators of a university, for instance how important is scientific collaboration, particularly specific types of collaboration? What is the role scientific quality measured by citation impact? Does the size of a university (in number of publications), or does the number of students matter? We composed a large database of city and university data: gross urban product and population data of nearly 200 German cities and 400 districts for the period 1997-2017. Data for the universities are derived from the CWTS bibliometric data system and supplemented with data on the number of students 1995-2020. Performance characteristics of universities are derived from the Leiden Ranking 2020. The socioeconomic strength of a city is determined with the urban scaling methodology. Our study shows a significant relation between the presence of a university in a city and its socioeconomic indicators, particularly for larger cities, and that this is especially the case for universities with higher values of their output, impact and collaboration indicators.*

# 1. Introduction

## 1.1 *Objective of the study*

Knowledge is the driving force of technological, socioeconomic and health-care innovations, and therefore a crucial source of sustainable wealth. Cities in particular are centers of knowledge production and innovation, especially when knowledge institutions such as a university are present. Concentration of knowledge production increases the attractiveness of cities for talented and entrepreneurial, highly skilled persons and thus these urban centers continuously reinforce their socioeconomic strength (Glaeser 1999, Bettencourt et al 2007a,b). There is an extensive literature on the relation between human capital and innovation at the one side, and increase of socioeconomic welfare at the other. Recent work focuses on the regional innovation impact of universities in Europe (Tijssen, Edwards, Jonkers 2021). Most of the literature is global and general in nature but there is little research on the economic impact of universities themselves on their urban and regional home base.

Recent research on a world-wide scale based on an analysis of 15,000 universities in about 1,500 regions in 78 countries shows that increases in the number of universities are positively associated with future growth of GDP per capita and that there appear to be positive spillover effects from universities to geographically close neighboring regions. This effect is not simply driven by direct expenditures of the university, its staff and students but also through an increased supply of human capital and greater innovation (Valero and van Reenen 2019). But also this study is rather general. There is a need for a study that looks closer at the direct socioeconomic effects of universities. In this study we investigate whether there is a significant relation between the mere presence of a university in a city, and a city's socioeconomic strength, the growth of its gross urban product, and its population size. And if so, what are the characteristics of a university that matter?

As in many countries, also in Germany almost all major cities (population >100,000) do have institutions of higher education, for example small, specialized colleges with the formal status of a university or other types of colleges. But only a small part of these higher education institutions can be characterized as major universities with a large research output of international level and a large number of students. In order to work with clear criteria on these aspects, the decision whether or not a city has a major university is based on the Leiden Ranking (Waltman et al 2012), version 2020.

The structure of this paper is as follows. First we discuss how we measure the socioeconomic strength of a city or district on the basis of the urban scaling methodology. We present results of these measurements for German cities in different regions of the country. The second part of the paper focuses on the socioeconomic position of university cities compared to other cities and which characteristics of universities play a significant role.

## 1.2 *What is Urban Scaling?*

Recent studies show a *more than proportional* (superlinear) increase of the socioeconomic performance of cities (measured by the gross urban product) in relation to population size (Bettencourt et al 2010, 2013; Lobo et al 2013). This *urban scaling* relation is described by a power-law dependence of the gross urban product on population size given by the relation

$$G(N) = aN^\beta \qquad \text{Eq.1}$$

where $G$ is the gross urban product[1] and $N$ the population size of a city. The exponent $\beta$ follows from the measurement; in most cases, values of the exponent are between 1.10 and 1.20. We refer to our recent work on urban scaling for further details (van Raan 2020). The urban scaling relation implies that a city twice as large (in population) as another city can be expected to have approximately a $2^{1.15}$ = 2.22 larger socioeconomic performance (in terms of the gross urban product). Urban scaling behavior is also found for human interactions in general and for knowledge production activities in cities (Schläpfer et al 2014; Arbesman et al 2009; Bettencourt et al 2007a; Nomaler et al 2014).

A simple way to understand this phenomenon is by seeing cities as a complex network. The larger the city in population size, the more network nodes. The nodes in the urban system are the inhabitants, social and cultural institutions, centers of education and research, firms, etcetera. The number of nodes has a *linear* dependence on size, but the links between nodes depend on size in a *superlinear* way. The links between these (clustered) nodes are crucial for new developments, reinforcement of urban facilities, and innovation. Because they increase superlinearly, the socioeconomic strength of cities increases more than proportionally with increasing population size.

In this paper we build on our recent empirical work on urban scaling of German cities (van Raan 2020) which implies that we use the term city *only* for cities defined as municipalities and not for the entire urban agglomerations such as the US metropolitan statistical areas (Bettencourt at al 2010) or the European functional urban agglomerations (Bettencourt and Lobo 2016; OECD 2019; Eurostat 2019) which consist of many independent municipalities that may or may not cooperate optimally.

## 2. Data, Analytical Method

For our analysis we apply the same approach as described in our recent paper on urban scaling and for the explanation we largely follow the relevant text in that paper (van Raan 2020). Germany with 83 million inhabitants consists of sixteen federal states. These federal states have a specific administrative structure in which cities and districts (Kreise) play a central role. In connection with the availability of data on the gross urban product (GUP) at the German Federal Statistical Bureau, we discuss this administrative structure in and around German cities in more detail. Most larger cities (above 100,000 inhabitants) are *kreisfrei* ('district-free'), i.e., cities of which the surrounding urban area belongs to the municipality of the city, and therefore we have in these cases a one-governance urban area (which is in fact the definition of the concept 'kreisfrei'). Germany currently has 107 *kreisfreie* cities, with a total population of about 27,000,000. *Kreise* are districts around mostly smaller cities consisting of between 10 and 50 municipalities; together the Kreise (in total 294) have about 56,000,000 inhabitants. In Kreise the administrative and economic centers are cities that are not-kreisfreie cities (because they formally belong to a Kreis) although they can be larger than smaller kreisfreie cities[2]. These central cities within a Kreis are called Kreis-city.

---

[1] Throughout the text of this paper we use the abbreviation GUP for the gross urban product. In the case of mathematical equations we use the shorter symbol *G*.

Most university cities are kreisfrei, but several German university cities (as far as included in the Leiden Ranking) are Kreis-cities. This in the case for Hanover, Aachen, Göttingen, Tübingen, Paderborn, Saarbrücken, Marburg, Giessen, Konstanz, Greifswald, and Freiberg (not to be confused with Freiburg). Although these these cities are similar to a kreisfreie city in every respect, for curious local political reasons they are not kreisfrei, and thus these cities belong to a specific Kreis (Göttingen, Tübingen, Paderborn, Marburg, Giessen, Konstanz, Greifswald, and Freiberg) or to an ad-hoc defined urban administrative district (Hanover, Aachen, Saarbrücken). This has consequences for the data collection. At the city level, the German Federal Bureau of Statistics has GUP data available for the 107 kreisfreie cities. For the not-kreisfreie cities GUP data are available at the level of the Kreis. Therefore, we collected for all kreisfreie cities and for all Kreise (period 1992-2017) data on the gross urban product (GUP). The German Federal Bureau of Statistics (2019) also provided data on the population size (number of inhabitants) for all cities and for all Kreise (period 1970-2019) and data on the number of students in all German universities and other higher education institutions (1992-2019).

In this study we characterize universities with a series of bibliometrc indicators. These indicators are calculated with the data from the Leiden Ranking 2020. We refer to the Leiden Ranking website (Leiden Ranking 2020) for details on the data collection, data analysis and calculation of the indicators, particularly the impact-indicators. In total 54 German universities are included in the Leiden Ranking. We consider these as Germany's major universities. These are the universities that meet the selection criterion for the Leiden Ranking: at least an annual average of 200 Web of Science indexed publications in the period 2015–2018. Only research articles and review articles published in international journals ('core publications') are taken into account. Other types of publications are not considered. The Leiden data include at least the major universities but, of course, the selection criterion is rather arbitrary. Indeed, just below the threshold of the selection criterion there are several other universities of considerable size in output, impact and student numbers. Nevertheless, all 72 Max Planck Institutes are located in or nearby universities covered by the Leiden Ranking.

In Table 1 we give an overview of the data sets of cities used in this study. For instance, there are in total 81 cities in Germany with more than 100,000 inhabitants, and 44 of these cities have a university included in the Leiden Ranking. Of these cities, 69 (84%) are kreisfrei, 39 of them have a university included in the Leiden Ranking. We remark that the 54 universities do not always correspond one-on-one to cites: large cities like Berlin, Munich and Hannover have more than one university included in the Leiden Ranking, and some universities are located in two cities such as Erlangen-Nurnberg and Duisburg-Essen. In total, the 54 universities relate to 51 cities[3].

---

[2] An example is Neuss (Nord Rhine-Westphalia) with about 155,000 inhabitants, but this city is not kreisfrei, it is the administrative center (Kreis-city) of the Rhein-Kreis Neuss which has a population of about 450,000. The Bavarian city Schwabach on the other hand with about 41,000 inhabitants is a kreisfreie city.

*Table 1. Overview of the kreisfreie and not-kreisfreie cities with more than 100,000 inhabitants and between 50,000 and 100,000 inhabitants. Numbers between square brackets indicate the number of cities with a university (as far as included in the Leiden Ranking). There are 16 kreisfreie cities with less than 50,000 inhabitants. There is one not-kreisfreie university city with less than 50,000 inhabitants.*

|  | kreisfrei | not kreisfrei | total |
|---|---|---|---|
| cities >100,000 | 68 [39] | 13 [5] | 81 [44] |
| cities 50,000-100,000 | 23 [1] | 87 [5] | 110 [6] |

## 3. Urban Scaling of German Cities

### 3.1 Scaling of the gross urban product

We refer to our recent publication (van Raan 2020) for an extensive analysis of urban scaling in the western, southern, middle, northern and eastern regions[4] of Germany and in the country as a whole. For this study, we updated the data up to and including 2017. Fig. 1 shows the results of the analysis, we compare the scaling of the southern cities to those of the other regions of Germany. In all cases the gross urban product (GUP) scales superlinearly with population, the scaling exponent ranges between and 1.03 and 1.34. As we clearly see, in urban scaling not only the power law exponent is a crucial parameter, but also the absolute difference in GUP between two sets of cities, i.e., the distance between the regression lines.

This difference in GUP is clearly visible in the lower right panel of Fig. 1 where we compare the urban scaling of the southern region to the eastern region. The southern cities are generally at a considerably higher GUP level as compared with the cities in the eastern region, which is the former DDR. But also the difference the between the wealthy southern region of Germany and the old industrial western region is striking. For an extensive discussion on this issue, explanations of how individual cities influence the measured scaling exponents, and for confidence intervals of the measured scaling exponents, we refer to van Raan (2020). Remarkably, if we join all regions and calculate the urban scaling of the entire country, we find a low superlinear scaling exponent 1.03, lower than most of the separate regions of the country, see Fig. 2. This phenomenon points at an often neglected issue in urban studies: the scaling of GUP with population size of cities in a country may heavily depend on the regional economy within the country. This has to be taken into account when the scaling exponent of an entire country is calculated.

---

[3] In the case of more than one university in a city, we characterize the city with the maximum indicator values of the universities, see footnote 9, Section 4.2. In the case of a university located in two cities, we assign the university to both cities.

[4] North Rhine-Westphalia, western region of Germany; Baden-Württemberg and Bavaria, southern region; Hesse, Rhineland-Palatinate, and Saarland, middle region; Bremen, Hamburg, Lower Saxony, and Schleswig-Holstein, northern region; Berlin, Brandenburg, Mecklenburg-Vorpommern, Saxony, Saxony-Anhalt and Thuringia, eastern region.

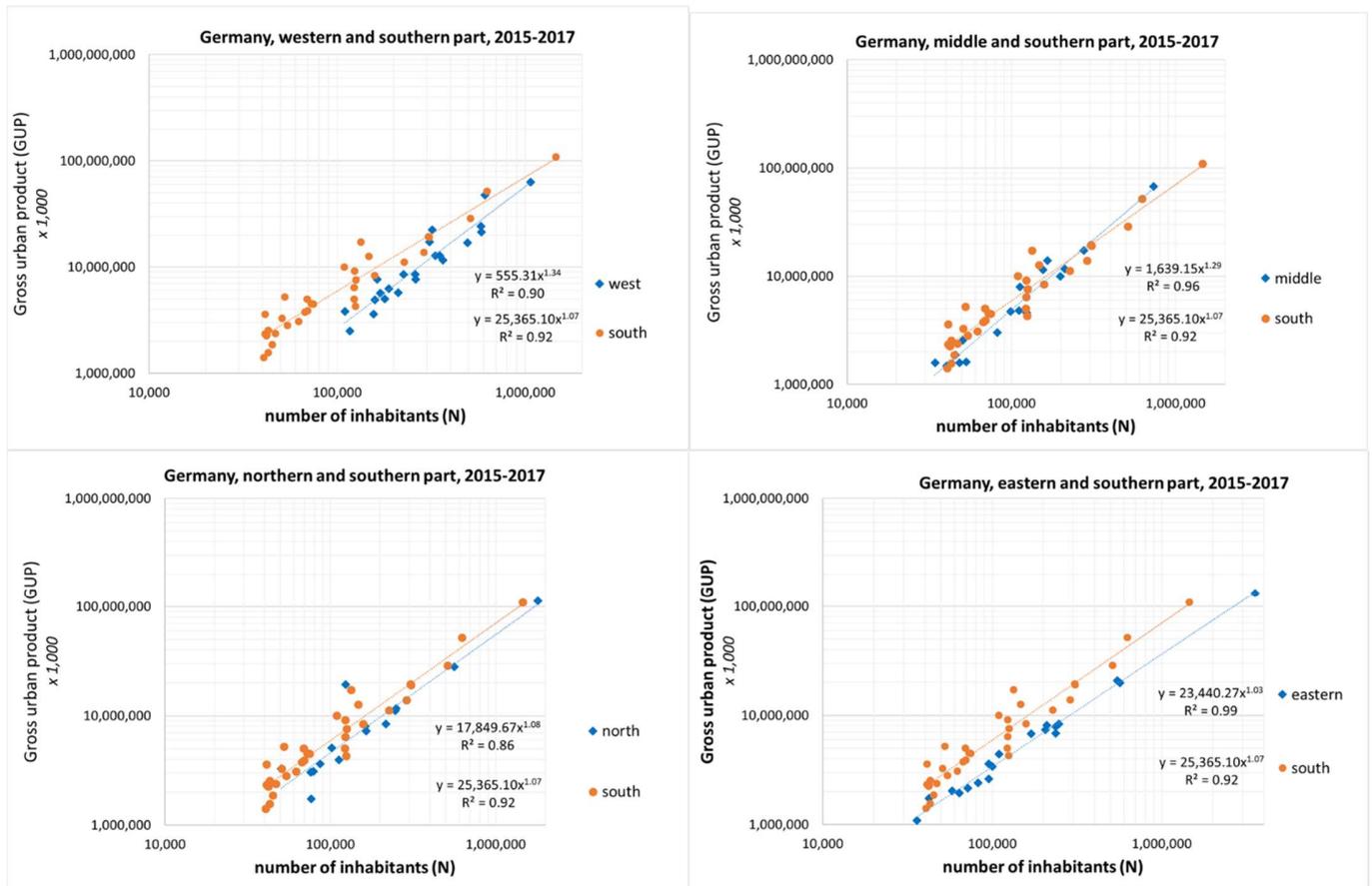

Fig. 1  Scaling of the gross urban product (GUP) for German (kreisfreise) cities. Upper left panel: western and southern region of Germany; upper right panel: middle and southern region; lower left panel: northern and southern region; lower right panel: eastern and southern region. (GUP in units of €1,000; data average 2015-2017).

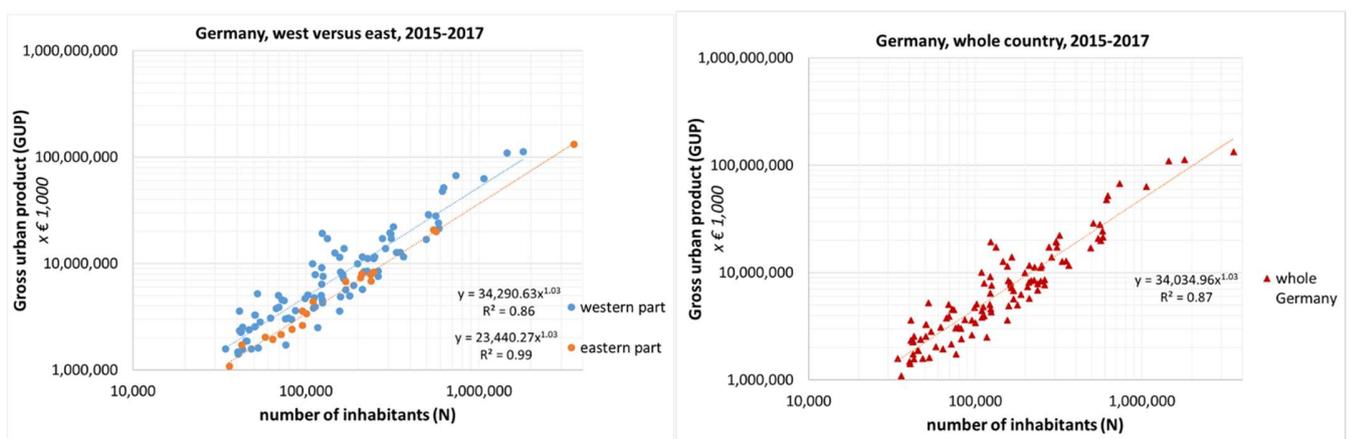

Fig. 2  Scaling of the gross urban product (GUP) for German (kreisfreise) cities. Left panel: all west regions versus east; right panel: all regions together, i.e., whole country (GUP in units of €1,000; data average 2015-2017).

### *3.2 Scaling Residuals as Indicator of Socioeconomic Strength*

As can be expected and also clearly visible in the empirical results (Figs. 1 and 2), the *observed* positions of cities will deviate from the *expected* positions given by the regression line through all measuring points of a specific set. These deviations can be measured by the residuals: Using Eq.1 (which is the scaling relation for a *set* of cities) we find that $G(N_i)$ is the *expected* gross urban product of an *individual* city ($i$) with population $N_i$. By denoting the *observed* (real) value of the gross urban product of a city with $G_i$, we calculate the residuals $\xi_i$ of the scaling distribution for each of the (kreisfreie) cities (and similar for the Kreise) as follows:

$$\xi_i = ln[G_i/G(N_i)] = ln[G_i/aN_i^\beta] \qquad \text{Eq.2}$$

Positive residuals indicate that a city performs better than expected. Thus, residuals can be used as an indicator of socioeconomic strength. Indeed, we find a strong correlation of residuals with other measures of socioeconomic strength, see van Raan (2020). Given the considerable economic differences between the different regions of Germany, we calculated the residuals both in relation to the *regional* ($\xi_r$) as well as to the *national* context ($\xi_n$). For instance, for cities in the western part of Germany the regional residuals are calculated with the relevant scaling law as presented in Fig. 1, upper left panel, i.e., on the basis of

$$G_{i,r}(N_i) = 555.31 N_i^{1.34}$$

whereas their national residuals are calculated with the scaling law in Fig. 2, lower panel, i.e., on the basis of

$$G_{i,n}(N_i) = 34034.96 N_i^{1.03}$$

Cities in the eastern part of Germany such as for instance Leipzig, Dresden and Jena do not yet have the same socioeconomic strength as many cities in other parts of Germany, but within their own region Leipzig, Dresden and Jena show a strong position. There is no rational basis on which to assign specific weights to the national and the regional residual as components in the calculation of an overall socioeconomic strength indicator. Therefore we give both components equal weight and take as a measure for the relative socioeconomic strength $S$ of a city the average value of the national and the regional residual:

$$S = (\xi_n + \xi_r)/2 \qquad \text{Eq.3}$$

The statistical uncertainty in this measure is determined by the uncertainty in the measured residuals, and these are determined by the standard error values of the measured scaling coefficient and scaling exponent. On the basis of our earlier discussion on confidence levels of scaling parameters (van Raan 2020) we estimate the uncertainty in $S$ to be $\pm 0.03$. An important characteristic of urban residuals is that they are quite stable and vary little over a long period, often timescales of several decades (Alves et al 2015; Bettencourt 2020). Thus, scaling residuals can be seen as reliable indicators of the socioeconomic strength of cities.

The full list for all kreisfreie cities with their national residuals $\xi_n$, regional residuals $\xi_r$ and socioeconomic strength $S$ is presented in Table S1. In Fig. S1 we show the normal distribution of these parameters calculated on the basis of their respective means and standard deviations. In Fig. S2 we show the rank-distribution of the national residual in comparison with the regional residual. In Table 2 we show as an example the top-25 cities ranked by their socioeconomic strength $S$.

*Table 2. German (kreisfreie) cities ranked by $S$, top-25 (university cities, as far as present in the Leiden Ranking, in bold).*

| kreisfreie city | $S$ |
|---|---|
| Wolfsburg | 1.20 |
| Ingolstadt | 0.90 |
| Schweinfurt | 0.68 |
| **Erlangen** | 0.56 |
| Coburg | 0.56 |
| Ludwigshafen | 0.49 |
| **Regensburg** | 0.48 |
| **Bonn** | 0.43 |
| **Düsseldorf** | 0.42 |
| Koblenz | 0.38 |
| Emden | 0.38 |
| **Darmstadt** | 0.37 |
| **Stuttgart** | 0.37 |
| **Frankfurt** | 0.35 |
| Aschaffenburg | 0.35 |
| **Ulm** | 0.34 |
| Passau | 0.25 |
| **München** | 0.24 |
| **Münster** | 0.21 |
| **Bayreuth** | 0.20 |
| Speyer | 0.18 |
| Memmingen | 0.17 |
| Leverkusen | 0.16 |
| Bamberg | 0.16 |
| Zweibrücken | 0.15 |

At the top of the $S$ ranking we see the cities Wolfburg and Ingolstadt with an extraordinary high socioeconomic strength. Wolfsburg (about 125,000 inhabitants) is the location of the Volkswagen (VW) headquarters with the world's biggest car plant, production of 815,000 cars per year (2015) and 70,000 employees in Wolfsburg alone. The city even owes its origins entirely to VW, founded in 1938 it had only 1,000 inhabitants. Now, measured in GUP per capita, Wolfsburg is one of the richest cities in Germany. Ingolstadt (about 140,000 inhabitants) is partly a similar case: this city is home to the headquarters of the automobile manufacturer Audi. But in strong contrast to Wolfsburg, Ingolstadt was already in the early Middle Ages an important city in Germany. Both cities are not cities with a major university, Wolfsburg has a college with a focus on vehice technology. Ingolstadt had a university from 1472-1800, and from 1980 this city has a small catholic university focusing mainly on social sciences and humanities (in this study not considered as a major university) as well as a technical college. As a result of the huge automobile companies both cities have an extraordinary large urban scaling residual.

We already noticed that not all university cities are kreisfreie cities. Therefore, they are not present in Tables S1 and 2. As discussed in Section 2, this is the case for Hanover, Aachen, Göttingen, Tübingen, Paderborn, Saarbrücken, Marburg, Giessen, Konstanz, Greifswald, and Freiberg. This means that no GUP data are directly avaliable for these cities, only for their total Kreis. Nevertheless, all the above mentioned cities are (often by far) the largest cities in the Kreis (and that is why they are called Kreis-city) and they will largely determine the socioeconomic position of their Kreis. Therefore we use the data of their Kreis to determine the socioeconomic strength of these cities. We calculated the Kreis residuals in the same way as we calculated the residuals of the kreisfreie cities, both in relation to the regional as well as to the national context, and determine the socioeconomic strength also in the same way as for the kreisfreie cities. In Table S2 we show the list of all Kreise with their national residuals $\xi_n$, regional residuals $\xi_r$ and socioeconomic strength $S$. Kreise with university cities (as far as these universities are present in the Leiden Ranking) are shown in Table 3.

Table 3.  German Kreise with university cities ranked by $S$.

| (not-kreisfreie) city | $S$ |
|---|---|
| **Saarbrücken** (Kreis Regionalverband Saarbrücken) | 0.23 |
| **Hannover** (Kreis Region Hannover) | 0.15 |
| **Paderborn** (Kreis Paderborn) | 0.10 |
| **Aachen** (Kreis Städteregion Aachen) | 0.09 |
| **Marburg** (Kreis Marburg-Biedenkopf) | 0.08 |
| **Göttingen** (Kreis Göttingen) | 0.06 |
| **Giessen** (Kreis Giessen) | 0.03 |
| **Tübingen** (Kreis Tübingen) | -0.02 |
| **Konstanz** (Kreis Konstanz) | -0.03 |
| **Freiberg** (Kreis Mittelsachsen) | -0.05 |
| **Greifswald** (Kreis Vorpommern-Greifswald) | -0.15 |

Using the residual calculations for the kreisfreie cities and for the Kreise as discussed above, we analyze in the next section all 191 German cities above 50,000 inhabitants, 50 of them are university cities, with a special focus on the cities above 100,000 inhabitants because the vast majority of university cities (44) are in this group.

# 4. University Performance and Socioeconomic Characteristics of their Cities

## 4.1 University cities compared to other cities

In this section we compare university cities to other cities on the basis of three different socioeconomic indicators: (1) the socioeconomic strength $S$; (2) the growth of the gross urban product over the last 20 years ($T$); and (3) the growth in population in the last 20 years ($U$). We make this comparison for cities between 50,000 and 100,000 inhabitants, and for cities above 100,000 inhabitants. We have calculated the socioeconomic strength $S$ in the foregoing section, in Fig. 4 we show the normal distribution of $S$ for cities > 100,000. In this figure, the first two quartiles of the distribution are marked.

We have GUP values for kreisfreie cities and Kreise available from 1992 to 2017 and calculate the ratio $T$ between the average GUP value for 2015-2017 and for 1995-1997:

$T$= [GUP(2015-2017)/GUP(1995-1997)].

This indicator defines the growth of the gross urban product over the last 20 years. We consider $T$ as the indicator of socioeconomic strengthening. We present this indicator for all cities between 50,000 and 100,000 and all cities above 100,000 inhabitants in Table S3. We see high $T$ values the car industry cities Ingolstadt and Wolfsburg. This is also the case for the former East German cities Jena, Potsdam, Dresden and Leipzig. It indicates the socioeconomic strengthening of these cities. Of the ten lowest ranked cities above 100,000 inhabitants the majority is in the old industrial Region (Ruhr Area) in Nord Rhine Westphalia. In Fig. 5 we present the normal distribution of $T$ for the 81 cities >100,000, also here the first two quartiles are marked.

Our third socioeconomic city indicator is the population growth. Using the data on city population we calculate the ratio $U$ between the number of inhabitants in 2019 and in 2000:

$U$ = [N(2019)/N(2000)].

We present this ratio for all cities between 50,000 and 100,000 and all cities above 100,000 inhabitants also in Table S3, where $S$ values and number of inhabitants $N$ are included. Some cities show a relatively strong growth, such as Potsdam while a considerable part of the cities (35%) did not grow at all or even decreased in population. In Fig. 6 we present the normal distribution of $U$ for the 81 cities >100,000, again with the first quartiles of the distribution indicated.

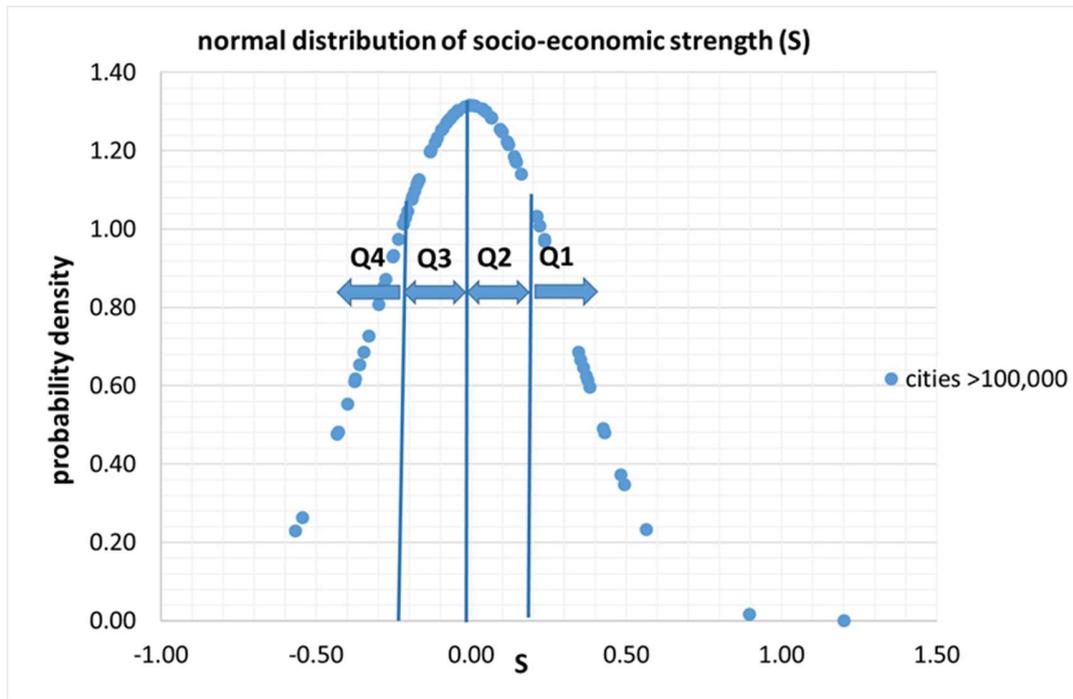

*Fig. 4  Normal distribution of the socioeconomic strength ($S$).*

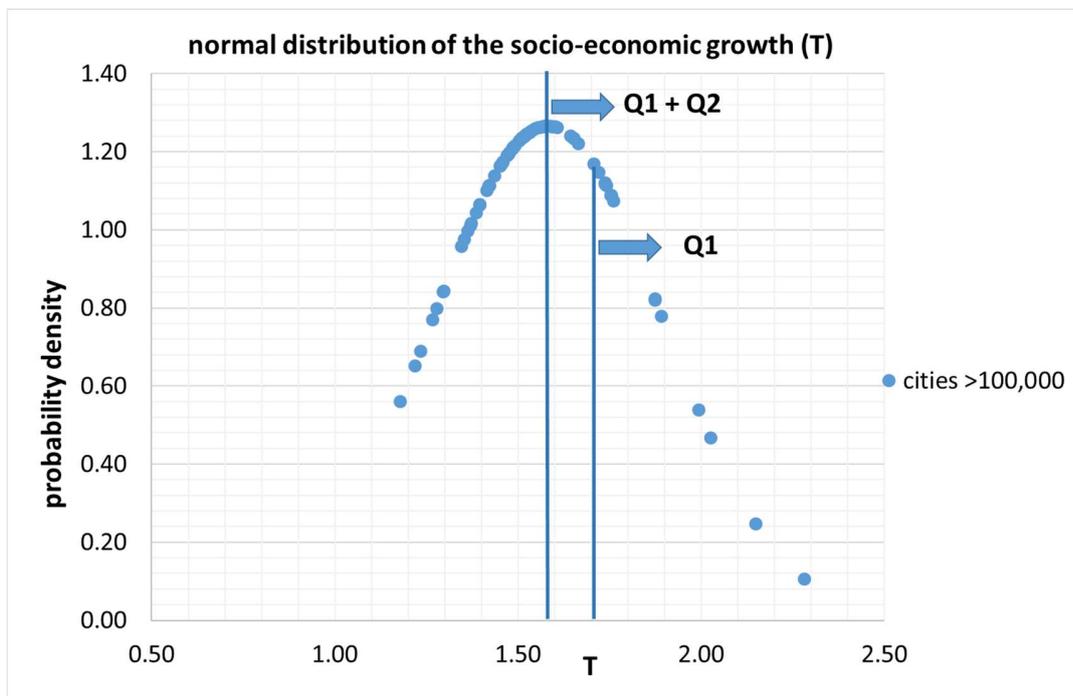

*Fig. 5  Normal distribution of the GUP increase over 20 years ($T$).*

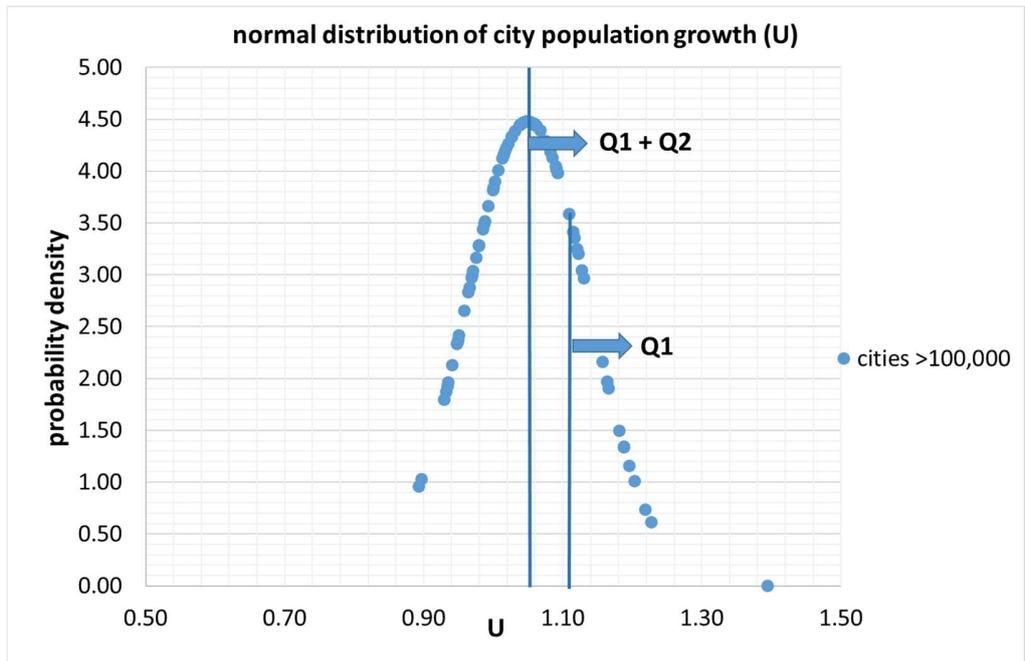

*Fig. 6 Normal distribution of the city population growth over 20 years ($U$).*

We analyze the data as follows and take the socioeconomic strength $S$ as an example. The *independent* variable concerns cities, namely *university cities* versus *other cities*. We rank all cities by their $S$ value and divide this ranking into quartiles in order to have a reasonable number of cities (20) per unit of division. Thus, the first quartile $S(Q1)$ are the cities in the top-25% of the $S$ distribution, and so on. For example, $S(Q1)$ of all cities >100,000 covers the values 1.20 to 0.15, the second quartile $S(Q2)$ the values 0.14 to -0.04, and so on, and in a similar way for the two other distributions, see Figs. 4-6 (the relevant data are in Table S3).

For each quartile we count the number of cities *with* universities as well cities *without* (Leiden Ranking) universities. The result of this analysis is presented in two contingency tables, see Table 4, left side for the 81 cities >100,000 and the right side for the 110 cities with 50,000 - 100,000 inhabitants. A chi-squared test of the data for $S$(Q1, Q2, Q3, and Q4) renders a p-value of 0.010 in the case of the cities >100,000 inhabitants. Taking the conventionally accepted significance level p<0.05 we find a significant difference in the distribution of cities according to the presence of universities.

A further inspection of the contingency table reveals more information. We see that it is particularly the fourth quartile $S(Q4)$ distribution that makes the difference: universities are significantly absent in cities with the lowest socioeconomic strength. By distinguishing between the first half of the $S$ distribution at the high values side, $S(Q1+Q2)$, and the low values second half $S(Q3+Q4)$, 28 of the 44 university cities >100,000 (64%) are in $S(Q1+Q2)$, which means that they are cities with an above-average socioeconomic strength (p=0.011, the probability that no difference exists). For the group of 110 cities with 50,000 – 100,000 inhabitants the significance disappears. But as is clear from Table 4, the number of university cities is this group is very low (6) and one cannot expect significant results.

*Table 4. Number of university cities and other cities for each quartile of the socioeconomic strength ($S$) distribution (Left side: the 81 cities above 100,000; right side the 110 cities 50,000 – 100,000)*

| $S$ | univ cities | other cities |     |
|-----|-------------|--------------|-----|
| Q1  | 12          | 8            | 20  |
| Q2  | 16          | 5            | 21  |
| Q3  | 11          | 9            | 20  |
| Q4  | 5           | 15           | 20  |
|     | 44          | 37           | 81  |
| p=  | 0.010       |              |     |

| $S$ | univ cities | other cities |     |
|-----|-------------|--------------|-----|
| Q1  | 1           | 26           | 27  |
| Q2  | 2           | 26           | 28  |
| Q3  | 2           | 25           | 27  |
| Q4  | 1           | 27           | 28  |
|     | 6           | 104          | 110 |
| p=  | 0.871       |              |     |

| $S$   | univ cities | other cities |    |
|-------|-------------|--------------|----|
| Q1+Q2 | 28          | 13           | 41 |
| Q3+Q4 | 16          | 24           | 40 |
|       | 44          | 37           | 81 |
| p=    | 0.011       |              |    |

| $S$   | univ cities | other cities |     |
|-------|-------------|--------------|-----|
| Q1+Q2 | 3           | 52           | 55  |
| Q3+Q4 | 3           | 52           | 55  |
|       | 6           | 104          | 110 |
| p=    | 1.000       |              |     |

We performed a similar analysis for the two other city indicators, the socioeconomic growth $T$ and the population growth $U$. The results are shown in Table S4. For the cities >100,000 we find that the university cities are significantly present in $T$(Q1+Q2). For the population growth $U$ we find that for the cities >100,000 as well as for the cities with 50,000-100,000 inhabitants the university cities are significantly present in $U$(Q1+Q2). Just like in the case of the socioeconomic strength $S$, the significance is mainly due to the low number of university cities in the fourth quartile. For the cities between 50,000 and 100,000 we also find that the university cities are significantly in $U$(Q1+Q2). Table 5 gives an overview of our findings. We conclude that for all three socioeconomic indicators (socioeconomic strength, socioeconomic growth, and population growth) university cities are predominantly present in the better half of these indicators.

*Table 5. Overview of the significance tests for university cities with respect to the three socioeconomic city indicators.*

|                         | population growth $U$ | socioeconomic strength $S$ | socioeconomic growth $T$ |
|-------------------------|-----------------------|----------------------------|--------------------------|
| *cities N 50,000-100,000* | significant           | not significant            | not significant          |
| *cities N>100,000*      | significant           | significant                | significant              |

So we find a positive relation between having a major university and being a city with a relatively strong socioeconomic position. However, we must be cautious with our conclusions. We certainly did not find an iron law for each university city, nor can we make a statement about causality. An indication of a possible causal relation is that most of the major universities are centuries old whereas our socioeconomic indicators relate to recent times. This temporal precedence could suggest that cities with a major university had a higher probability than cities without a major university to develop into a socioeconomically strong position.

Next to the data on socioeconomic strength, socioeconomic growth and population growth used so far in the study, there is a further important source of data. The German socioeconomic research agency Prognos AG (Prognos 2021) evaluates the future opportunities and risks of all (kreisfreie) cities and Kreise. Prognos publishes the results every three years since 2004 in the report *Zukunftatlas*. The last edition is from 2019. The evaluation of the future perspectives of cities and Kreise is based on 29 macro- and socioeconomic indicators to assess strength and dynamism. These indicators cover the fields of demographics, labor market, social welfare, competition and innovation. No scaling approaches were applied. These indicators are discussed in detail in the Zukunftatlas (Prognos Zukunftatlas 2019). On the basis of these assessments, a ranking (Future Index) of all cities and Kreise is created. The Prognos Future Index is the only nationwide German ranking that shows urban regional developments over a period of 15 years. We used the publicly available[5] overview of all rankings since 2004 to calculate for all German cities with a population larger than 100,000 the difference in Prognos ranking positions between the years 2004 and 2019. For instance, the Prognos ranking position of Berlin was 262 in 2004, and 93 in 2019. So Berlin improved its ranking position with +169. In sharp contrast, the traditional old industry city Essen falls in ranking positon from 121 in 2004 to 239 in 2019, a difference of -131.

After distinguishing between university cities and other cities we calculated for both groups the normal distribution of ranking position differences. Fig. S3 presents the results. We tested the difference between the two means of both distributions and found that at the 95% confidence interval level p=0.041. We conclude that university cities improved their ranking positions in the 15 years period between 2004 and 2019 significantly more than the other cities.

### 4.2   *Bibliometric Performance Indicators of German Universities*

In the previous section we compared university cities with other cities. We found that university cities are predominantly present in the better half of the $S$, $T$ and $U$ distributions but not all cities with a major universities belong to the socioeconomically strong cities. Could this be due to characteristic differences between universities? In other words: how do the university cities in the first quartile or first two quartiles of the $S$, $T$ and $U$ distributions differ from the university cities in the second, third and fourth quartile, or in the third and fourth quartile, respectively? In order to investigate this, we have to categorize the university cities in such a way that we can distinguish them from one another. We do this with help of bibliometric[6] indicators. Before we perform the analysis, we first have to discuss the basic elements of bibliometric indicators.

---

[5] https://de.wikipedia.org/wiki/Zukunftsatlas. We checked the reliability of the data in this Wikipedia page with the original data in the Prognos Zukunftatlas 2016 and 2019.

[6] The quantitative study of science, mostly referred to as *scientometrics*, aims at the advancement of our knowledge on the development of science and its communication structure, in relation to social,

Distinguishing between universities is the core business of university rankings. On the basis of survey data or bibliometric data, or both, several organizations produce annual rankings of universities. Frequently used rankings are the Academic World Universities Ranking ("Shanghai Ranking", ARWU 2020), the Times Higher Education (THE 2020) ranking, the Leiden Ranking (Leiden Ranking 2020), the QS ranking (QS 2020), the Scimago ranking (Scimago 2020) and the U-Multirank (U-Multirank 2020). For an extensive discussion of the problems related to university rankings, we refer to van Raan (2005, 2019) and Waltman et al (2012). Here we briefly outline several important issues. The combination of scores for teaching and research performance to one final score is methodologically incorrect because teaching and research are different tasks and also different missions of universities. In research rankings it is incorrect to combine size-dependent (e.g., number of papers in specific journals such as Nature and Science) and size-independent measures (e.g., publications per staff member). Indicators based on citation analysis must be field-normalized, otherwise universities with a focus on engineering, or on social sciences and humanities will be systematically disadvantaged. Often citation indicators are based on averages. But average-based indicators are very sensitive to outliers, thus they are not the best statistic in the case of skewed distributions, such as the distribution of citations over publications. A further problem is that comparison of ranking scores in a time series can be affected seriously if meanwhile the number of universities covered by the ranking is increased substantially, for instance by lowering the required threshold for the number of publications. Last but not least, the definition of a university, particularly the relation with medical schools and hospitals, is a cumbersome task.

All above issues, including a well-defined uncertainty measure, are dealt with meticulously in the Leiden Ranking (Waltman et al 2012). For this paper we use the 2020 version of the Leiden Ranking (Leiden Ranking 2020). In this version publication data relate to the time period 2015-2018 and the citation data to 2015-2019, author self-citations are excluded. Universities are included if they have more than 200 publications covered by the Web of Science (WoS)[7] on average per year in the period 2015-2018. We consider this also as the definition for a major university. This does not mean that universities with a publication output below the above mentioned threshold are low performance institutions but they are not a major university in terms of scientific productivity.

We consider in our analysis the indicators given in Table 6. In the Leiden Ranking publications and citations can be *fractionally* or *fully* counted[8]. We distinguish 10 indicator families

---

technological and socioeconomic aspects. Within scientometrics, research on scientific communication, particularly with data from publications, citations, and journals is called *bibliometrics* (van Raan 2019).

[7] Web of Science, published by Clarivate Analytics, see https://clarivate.com/webofsciencegroup.

[8] The scientific impact indicators in the Leiden Ranking are calculated using either a full or a fractional counting method. The full counting method gives equal weight (with value 1) to all publications of a university, regardless of collaboration. The same goes for the citations received by these publications. The fractional counting method, however, gives less weight to collaborative publications than to non-collaborative ones. More specifically, publications as well as their citations are divided over the collaborating institutes. The fractional counting method leads to a more proper field normalization of impact indicators (Waltman and van Eck 2015). Because of the better normalization properties, fractional counting is regarded as the preferred method in the Leiden Ranking, but both modalities are available in this ranking. The advantage in having both is that it provides a good idea of the robustness of the outcomes. At high aggregation levels such as universities, the correlation between the ranking based on full counting and that based on fractional counting is high.

consisting of one to at most 4 sub-indicators. For instance, the second indicator family contains 4 sub-indicators: the (absolute) number of fractionally counted publications in the worldwide top-1, 5, 10, 50% of the citation-impact distribution of the relevant field (*Pt1frac, Pt5frac, Pt10frac, Pt50frac*). The first nine indicator families are available in the Leiden Ranking, the tenth (number of students, year 2020, and the increase of the number of students in the last 20 years) was obtained from the German Federal Bureau of Statistics. In total we have 26 (sub)indicators which means that each university (*i*) is characterized by the set of indicators {$i1, i2, …, i26$}. Consequently, also the university city is characterized by these indicators[9].

The *p(t10frac)* indicator (third indicator family, third sub-indicator) is generally considered as the main research performance indicator. This indicator gives the fraction of publications that are in the top-10% of their fields[10] in the case that publications are fractionally counted. So if for a university this fraction is 0.100, this university performs according to the expected value; if the fraction is above 0.100, the university performs better, and below 0.100 the performance is lower than the expected value. The indicator values of the universities have in good approximation a normal distribution. As an example we show in Fig. S4 this distribution of the *p(t10frac)* indicator for all German universities covered by the Leiden Ranking.

*Table 6. University indicators considered in this study.*

| 1 | Number of fractionally counted publications (*Pfrac*); |
|---|---|
| 2 | Number of fractionally counted publications in the top-1, 5, 10, 50% (*Pt1frac, Pt5frac, Pt10frac, Pt50frac*) |
| 3 | Same as 2, now relative (*pt1frac*=[*Pt1frac/Pfrac*]), similar for *pt5frac, pt10frac, pt50frac*) |
| 4 | Number of fully counted publications (*Pfull*) |
| 5 | Number of fully counted publications in the top-1, 5, 10, 50% (*Pt1full, Pt5full, Pt10full, Pt50full*) |
| 6 | Same as 5 now relative (*pt1full*=[*Pt1full/Pfull*]), similar for *pt5full, pt10full, pt50full*) |
| 7 | Number of fractionally counted citations (*Cfrac*) |
| 8 | Number of fully counted citations (*Cfull*) |
| 9 | Number of collaborative publications (total *Pcoll*, within this total: international *Pintcoll*, with business companies *Pb*, and of these latter with local business companies *PbL*) (fully counted) |
| 10 | Number of students (*Ns*) and the increase of this number in the last 20 years (*V*) |

In Table 7 we present German universities cities (as far as included in the Leiden Ranking 2020) ranked by the *pt10frac* indicator of their university (in case of more than one university, see footnote 10). We show the first 25 cities (two of which have a population below 100,000) and given the large amount of data we limit the table to the first nine indicators (indicator families 1 to 3) and the last two indicators (indicator family 10), as well as the population of the city ($N$, year 2019) and the values of the three socioeconomic indicators $S$, $T$ and $U$. The complete set of data (all university cities, all indicators) is available in our data repository[11].

---

[9] In the case of for instance two universities *a* and *b* in one city, we characterize that city as if it has one university with the set of indicators {max(*a*1,*b*1), max(*a*2,*b*2),……, max(*a*26,*b*26)}.

[10] We use here the indictor symbol *p(t10frac)*, in the Leiden Ranking this indicator has the symbol PP(top 10%) calculated in the fractional counting modality.

[11] See https://osf.io/4ru96/.

An illustration of the differences in student numbers ($Ns$) and growth in student numbers ($V$) for universities in the Leiden Ranking (LR) and universities/other higher education institutions not LR is given in Fig. S5. We find a significant difference in student numbers between the top-universities (i.e., LR universities in the first quartile of the *pt10frac* distribution) and all LR universities (p=0.012), and also a significant difference between all LR universities and the not-LR universities (p<0.000). For the growth in student number there is no significant difference between the top-universities and all LR universities, but a significant difference between all LR universities and the not-LR (p=0.002): not-LR universities show a larger increase of student numbers as compared to the LR universities.

*Table 7. German university cities with the first three bibliometric indicator families (9 indicators) and indicator family 10 (number of students and growth of the number of students, main text), as well as city population ($N$) and the three socioeconomic indicators $S$, $T$, and $U$, ranked by the p(t10frac) indicator (we show the first 25).*

| Univ city | $N$ | $N(s)$ | $S$ | $U$ | $T$ | $V$ | Pfrac | Pt1frac | Pt5frac | Pt10frac | Pt50frac | pt1frac | pt5frac | pt10frac | pt50frac |
|---|---|---|---|---|---|---|---|---|---|---|---|---|---|---|---|
| Göttingen | 118,911 | 30162 | 0.06 | 0.96 | 1.51 | 1.32 | 4872 | 60 | 346 | 664 | 2838 | 0.012 | 0.071 | 0.136 | 0.582 |
| München | 1,484,226 | 48697 | 0.24 | 1.23 | 1.75 | 1.53 | 8142 | 101 | 498 | 1000 | 4681 | 0.014 | 0.066 | 0.133 | 0.575 |
| Bonn | 329,673 | 38481 | 0.43 | 1.09 | 1.42 | 1.04 | 4819 | 70 | 287 | 590 | 2734 | 0.015 | 0.060 | 0.122 | 0.567 |
| Heidelberg | 161,485 | 25986 | -0.01 | 1.15 | 1.74 | 1.26 | 7744 | 100 | 492 | 946 | 4374 | 0.013 | 0.064 | 0.122 | 0.565 |
| Würzburg | 127,934 | 27552 | 0.14 | 1.00 | 1.52 | 1.68 | 3622 | 45 | 215 | 441 | 2020 | 0.012 | 0.059 | 0.122 | 0.558 |
| Münster | 315,293 | 45022 | 0.21 | 1.19 | 1.53 | 1.03 | 4707 | 52 | 309 | 565 | 2579 | 0.011 | 0.066 | 0.120 | 0.548 |
| Mainz | 218,578 | 29907 | 0.04 | 1.20 | 1.46 | 1.11 | 3817 | 48 | 229 | 443 | 2098 | 0.012 | 0.060 | 0.116 | 0.550 |
| Stuttgart | 635,911 | 24153 | 0.37 | 1.09 | 1.60 | 1.71 | 2697 | 29 | 153 | 310 | 1499 | 0.011 | 0.057 | 0.115 | 0.556 |
| Frankfurt | 763,380 | 45179 | 0.35 | 1.18 | 1.50 | 1.26 | 4462 | 51 | 264 | 510 | 2443 | 0.011 | 0.059 | 0.114 | 0.548 |
| Erlangen | 112,528 | 37575 | 0.56 | 1.12 | 2.02 | 1.92 | 5939 | 74 | 351 | 678 | 3198 | 0.013 | 0.059 | 0.114 | 0.538 |
| Nürnberg | 518,370 | 37575 | 0.00 | 1.06 | 1.65 | 1.92 | 5939 | 74 | 351 | 678 | 3198 | 0.013 | 0.059 | 0.114 | 0.538 |
| Freiburg | 231,195 | 24028 | -0.10 | 1.13 | 1.74 | 1.37 | 4923 | 56 | 279 | 561 | 2809 | 0.011 | 0.057 | 0.114 | 0.570 |
| Aachen | 248,960 | 45945 | 0.09 | 1.02 | 1.60 | 1.58 | 6146 | 64 | 344 | 694 | 3280 | 0.010 | 0.056 | 0.113 | 0.534 |
| Karlsruhe | 312,060 | 23616 | 0.12 | 1.12 | 1.54 | 1.73 | 5527 | 55 | 313 | 618 | 3021 | 0.010 | 0.057 | 0.112 | 0.547 |
| Berlin | 3,669,491 | 37312 | -0.17 | 1.08 | 1.57 | 1.05 | 5284 | 57 | 278 | 570 | 2868 | 0.011 | 0.057 | 0.111 | 0.548 |
| Köln | 1,087,863 | 54105 | 0.05 | 1.13 | 1.57 | 0.91 | 4029 | 39 | 217 | 447 | 2177 | 0.010 | 0.054 | 0.111 | 0.540 |
| Regensburg | 153,094 | 20584 | 0.48 | 1.22 | 1.99 | 1.44 | 2856 | 29 | 157 | 314 | 1576 | 0.010 | 0.055 | 0.110 | 0.552 |
| Essen | 582,760 | 43029 | -0.19 | 0.98 | 1.37 | 1.15 | 3424 | 32 | 186 | 375 | 1835 | 0.009 | 0.054 | 0.109 | 0.536 |
| Duisburg | 498,686 | 43029 | -0.36 | 0.97 | 1.43 | 1.15 | 3424 | 32 | 186 | 375 | 1835 | 0.009 | 0.054 | 0.109 | 0.536 |
| Darmstadt | 159,878 | 25170 | 0.37 | 1.16 | 1.59 | 1.57 | 2517 | 23 | 129 | 274 | 1349 | 0.009 | 0.051 | 0.109 | 0.536 |
| Kassel | 202,137 | 22786 | -0.05 | 1.04 | 1.45 | 1.49 | 863 | 7 | 45 | 92 | 418 | 0.009 | 0.053 | 0.106 | 0.484 |
| Dresden | 556,780 | 29148 | -0.09 | 1.17 | 1.87 | 1.28 | 4933 | 45 | 262 | 520 | 2584 | 0.009 | 0.053 | 0.105 | 0.524 |
| Bayreuth | 74,783 | 12931 | 0.20 | 1.01 | 1.60 | 1.85 | 1629 | 13 | 86 | 171 | 851 | 0.008 | 0.053 | 0.105 | 0.522 |
| Kiel | 246,794 | 27101 | -0.07 | 1.06 | 1.47 | 1.35 | 3087 | 40 | 161 | 324 | 1639 | 0.013 | 0.052 | 0.105 | 0.531 |
| Tübingen | 91,506 | 26842 | -0.02 | 1.13 | 1.90 | 1.45 | 5148 | 47 | 254 | 539 | 2776 | 0.009 | 0.049 | 0.105 | 0.539 |

## 4.3 University Performance and Socioeconomic Indicators of Cities

We are now ready to address the question *how* the university cities (>100,000 inhabitants) in the first quartile (Q1) or first two quartiles (Q1+Q2) of the $S$, $T$ and $U$ distributions differ from the other university cities. The $S$, $T$ and $U$ distributions are based on *all* 81 cities >100,000.. We apply two data-analytical methods. In the first method *the city indicators* are leading whereas in the second method *the university indicators* are leading.

We start with the first method. University cities are ranked by a specific city indicator (we do this successively for $S$, $T$ and $U$). As an example we take the $S$ distribution. For the university cities in the first quartile $S$(Q1) as well as for those in the other quartiles $S$(Q2+Q3+Q4) we calculate the mean and standard deviation of all university indicators and of the city indicators as well. With a test of the difference between the means we are able to find which indicators differ significantly when comparing the university cities in $S$(Q1) with those in $S$(Q2+Q3+Q4). We repeat the same procedure for the university cities in $S$(Q1+Q2) (above average socioeconomic strength) versus those in $S$(Q3+Q4) (below average socioeconomic strength). This analysis answers the question: are the university cities in the 'top' of a specific city indicator also the cities that have (on average) a significantly higher score for one or more university indicators, and which indicators are they?

The results of method 1 are presented in Table 8. The basic data and the calculations of the statistical significance are available in our data repository. We first give an example how to read this table. The left part of Table 8 relates to the university cities within $U$(Q1), the first quartile of the population growth distribution $U$ of all 81 cities >100,000. We find that for these cities the marked indicators have significantly larger vales as compared to the universities cities in the rest of the $U$ distribution, i.e., in $U$(Q2+Q3 +Q4). This difference is given by the ratio in the second column, and the p value in the third column gives the probability within the 95% confidence interval. Thus, for university cities in $U$(Q1) the *pt10frac* indicator value of their universities is 1.12 larger than the same indicator for the universities of the cities in $U$(Q2+Q3 +Q4), with p=0.009.

We conclude from Table 8 that universities in cities with an above-average population growth are in general universities with a higher performance in scientific output (publication-based indicators), in scientific impact (citation-based indicators) and in scientific collaboration. We also see in Table 8 that particularly the number of publications with local companies, i.e., companies in these cities and in their urban region (indicator *PbL*) is almost a factor two higher (1.72 in the case of $U$(Q1), p=0.048; and 2.08 in the case of $U$(Q1+Q2), p=0.029). Given that the number of publications of a university correlates quite well (van Raan 2006) with the size of the academic research staff, the significantly higher scores for the absolute number of publications *Pfrac* and *Pfull* suggest that the size of the staff, which can be regarded as a pool of innovative people, could be a significant parameter in relation to the population growth of the city.

The right part of Table 8 shows our findings with city indicator $S$, the distribution of the socioeconomic strength of cities. In this case we find that less university indicators than in the case of population growth correlate with the socioeconomic strength of a city. However, particularly the universities in cities in the first two quartiles $S$(Q1+Q2) show a higher performance as compared to the universities in cities in $S$(Q3+Q4) for, remarkably, especially the *fractionally* counted top 1, 5, 10 and 50 percent impact indicators, both in absolute (e.g., *Pt10frac*, p=0.017) as well as in relative terms (e.g., *pt10frac*, p=0.003). Particularly these fractionally counted relative top impact indicators are a strong indicator of scientific quality. These findings suggest that for university cities with an above-average socioeconomic strength the probability that their university is a top-university, is higher as compared to cities in the below-average socioeconomic strength.

Table S5 presents our findings for the third city indicator $T$, the growth in socioeconomic strength. Remarkably, for university cities in $T$(Q1) as well as in $T$(Q1+Q2) we do not find any university indicator that scores significantly higher as compared to the cities in $T$(Q2+Q3+Q4) and $T$(Q3+Q4), respectively. We do see however that socioeconomic growth and city population growth correlate significantly. This is to be expected, given the urban scaling relation between city population and the gross urban product of a city. We find that cities in the first quartile of the socioeconomic growth distribution have a 1.09 larger population growth as compared to the other cities (p=0.001).

Also the student population of a city can be regarded as a pool of innovative people. So, does the number of students ($Ns$, not included in Table 8) relates to one or more city indicators? We do not find a significant relation between number of students and city indicators. This does not mean that size of student population does not matter: our group of universities consists of mostly large universities with high numbers of students, the average number of students is around 30,000. Apparently within that order of magnitude, further differences in student population do not give significant correlations with city indicators.

Table 8. The marked indicators are significantly larger for the first quartile (Q1) or first half (Q1+Q2) of the $U$ and $S$ distributions for the 81 cities >100,000 as compared with the rest of this distribution. In the case of the distribution we do not have a ratio but a difference.

| U(Q1) | Ratio or diff: Q1/(Q2+Q3+Q4) | p | U(Q1+Q2) | Ratio or diff: (Q1+Q2)/(Q3+Q4) | p | S(Q1) | Ratio or diff: Q1/(Q2+Q3+Q4) | p | S(Q1+Q2) | Ratio or diff: (Q1+Q2)/(Q3+Q4) | p |
|---|---|---|---|---|---|---|---|---|---|---|---|
| $S$ | 0.15 | 0.037 | $S$ | 0.25 | 0.000 | $S$ | 0.43 | 0.0000 | $S$ | 0.37 | 0.000 |
| $U$ | 1.15 | 0.000 | $U$ | 1.14 | 0.000 | $U$ | | | $U$ | 1.06 | 0.017 |
| $T$ | 1.15 | 0.000 | $T$ | 1.14 | 0.000 | $T$ | | | $T$ | 1.08 | 0.045 |
| $V$ | | | $V$ | | | $V$ | | | $V$ | | |
| Pfrac | 1.59 | 0.002 | Pfrac | 1.56 | 0.011 | Pfrac | | | Pfrac | 1.44 | 0.031 |
| Pt1frac | 1.79 | 0.003 | Pt1frac | 1.82 | 0.009 | Pt1frac | | | Pt1frac | 1.70 | 0.017 |
| Pt5frac | 1.78 | 0.001 | Pt5frac | 1.68 | 0.015 | Pt5frac | | | Pt5frac | 1.67 | 0.014 |
| Pt10frac | 1.71 | 0.002 | Pt10frac | 1.63 | 0.017 | Pt10frac | | | Pt10frac | 1.61 | 0.017 |
| Pt50frac | 1.63 | 0.003 | Pt50frac | 1.60 | 0.011 | Pt50frac | | | Pt50frac | 1.51 | 0.024 |
| pt1frac | 1.18 | 0.030 | pt1frac | 1.20 | 0.018 | pt1frac | | | pt1frac | 1.21 | 0.012 |
| pt5frac | 1.16 | 0.005 | pt5frac | 1.13 | 0.027 | pt5frac | | | pt5frac | 1.17 | 0.004 |
| pt10frac | 1.12 | 0.009 | pt10frac | | | pt10frac | | | pt10frac | 1.13 | 0.003 |
| pt50frac | 1.04 | 0.021 | pt50frac | | 0.012 | pt50frac | 1.04 | 0.046 | pt50frac | 1.05 | 0.002 |
| Pfull | 1.58 | 0.006 | Pfull | 1.66 | 0.008 | Pfull | | | Pfull | | |
| Pt1full | 1.69 | 0.015 | Pt1full | 1.88 | 0.013 | Pt1full | | | Pt1full | | |
| Pt5full | 1.69 | 0.008 | Pt5full | 1.80 | 0.011 | Pt5full | | | Pt5full | | |
| Pt10full | 1.67 | 0.008 | Pt10full | 1.76 | 0.011 | Pt10full | | | Pt10full | | |
| Pt50full | 1.61 | 0.007 | Pt50full | 1.71 | 0.008 | Pt50full | | | Pt50full | | |
| pt1full | | | pt1full | | | pt1full | | | pt1full | | |
| pt5full | | | pt5full | | | pt5full | | | pt5full | | |
| pt10full | 1.09 | 0.047 | pt10full | | | pt10full | | | pt10full | | |
| pt50full | 1.04 | 0.027 | pt50full | | 0.021 | pt50full | | | pt50full | | |
| Cfrac | 1.67 | 0.005 | Cfrac | 1.57 | 0.031 | Cfrac | | | Cfrac | 1.63 | 0.017 |
| Cfull | 1.64 | 0.016 | Cfull | 1.73 | 0.021 | Cfull | | | Cfull | | |
| Pcoll | 1.59 | 0.009 | Pcoll | 1.71 | 0.008 | Pcoll | | | Pcoll | | |
| Pintcoll | 1.65 | 0.005 | Pintcoll | 1.75 | 0.007 | Pintcoll | | | Pintcoll | | |
| Pb | 1.61 | 0.018 | Pb | 1.80 | 0.012 | Pb | | | Pb | | |
| PbL | 1.73 | 0.048 | PbL | 2.08 | 0.029 | PbL | | | PbL | | |

In the second method *the university indicators are leading*: the university cities are ranked by a specific university indicator and we do this successively for all 26 university indicators. The data-analytical and statistical procedure are the same as in the first method. As an example we take the *p(t10frac)* distribution. For the university cities in the first quartile *p(t10frac)*(Q1) as well as for those in the other quartiles *p(t10frac)*(Q2+Q3+Q4) we calculate the mean and standard deviation of all university indicators and of the city indicators as well. With a test of the difference between the means we are able to find which indicators differ significantly when comparing the university cities in *p(t10frac)*(Q1) with those in *p(t10frac)*(Q2+Q3+Q4). We repeat the same procedure for the university cities in *p(t10frac)*(Q1+Q2) versus those in $S$ *p(t10frac)*(Q3+Q4). This analysis answers the question: are the university cities that are 'top' in a specific university indicator, for instance *p(t10frac)*, also the cities with a significantly higher $S$, $T$ and $U$ ?

Table 9 presents the results of method 2. We again see that the indicator *PbL*, the *absolute* number of scientific collaboration publications with local companies, has a significant relation with city indicator for population growth $U$: university cities in the first quartile *PbL*(Q1) have a significantly (p=0.019) larger city population growth as compared to the university cities in *PbL*(Q2+Q3+Q4). A further analysis shows that also the *relative* number of scientific collaboration papers in general *pcoll* and particularly for international collaborations *pintcoll* relate to city growth.

For the socioeconomic strength indicator $S$ we find that cities of which the university is in the first quartile of the *p(t10frac)* distribution has, on average, a significantly larger socioeconomic strength indicator value $S$ (with p=0.002). This is also the case for the other relative numbers of fractionally counted publications in the top-1, 5, and 10%. This is largely similar to what we found with method 1 although there also the absolute numbers of fractionally counted top-publications showed a significant relation. Whereas with method 1 not any university indicator appeared to relate significantly with the socioeconomic growth $T$ of the university cities, with method 2 we find a significant relation for *relative* number of scientific collaboration papers in general *pcoll*.

We cannot expect that all the results are the same for both methods: in method 1 city indicators are leading and thus their quartiles are based on the entire city indicator distribution for the 81 cities >100,000, whereas in method 2 the university indicators quartiles are based on the university indicator distributions of the 44 university cities. We illustrate the effect of this difference with an example in Table S6.

*Table 9. University indicators with a significant relation with city indicators S, U, and T.*

| pt1frac(Q1) | Ratio: Q1/(Q2+Q3+Q4) | p |
|---|---|---|
| S | 0.17 | 0.041 |

| pt5frac(Q1) | Ratio or diff: Q1/(Q2+Q3+Q4) | p |
|---|---|---|
| S | 0.18 | 0.027 |

| pt10frac(Q1) | Ratio or diff: Q1/(Q2+Q3+Q4) | p | pt10frac(Q1+Q2) | Ratio or diff: (Q1+Q2)/(Q3+Q4) | p |
|---|---|---|---|---|---|
| S | 0.25 | 0.002 | S | 0.16 | 0.021 |

| pt50frac(Q1+Q2) | Ratio or diff: (Q1+Q2)/(Q3+Q4) | p |
|---|---|---|
| S | 0.17 | 0.014 |

| pt50full(Q1+Q2) | Ratio or diff: (Q1+Q2)/(Q3+Q4) | p |
|---|---|---|
| S | 0.15 | 0.025 |

| Pt5frac(Q1) | Ratio or diff: Q1/(Q2+Q3+Q4) | p |
|---|---|---|
| S | 0.17 | 0.043 |

| PbL(Q1) | Ratio or diff: Q1/(Q2+Q3+Q4) | p |
|---|---|---|
| U | 1.07 | 0.019 |

| pcoll(Q1) | Ratio or diff: Q1/(Q2+Q3+Q4) | p |
|---|---|---|
| U | 1.08 | 0.006 |
| T | 1.09 | 0.026 |

| pintcoll(Q1) | Ratio or diff: Q1/(Q2+Q3+Q4) | p | pintcoll(Q1+Q2) | Ratio or diff: Q1/(Q2+Q3+Q4) | p |
|---|---|---|---|---|---|
| U | 1.06 | 0.045 | U | 1.06 | 0.014 |

We also investigated how the Prognos Future Index relates to our city indicators $S$, $T$ and $U$. To this end we used the ranking for all cities and Kreise of 2019 from which we deduced the ranking of the 81 cities >100,000. Next, we determined the quartiles of the distribution of this ranking ($P$ distribution, in which rank 1 is given the value 100, and so on). Then we calculated for $P$(Q1) of the 44 university cities the average values of $S$, $T$ and $U$ and compared these values with those for $P$(Q2+Q3+Q4). We did a similar comparison for $P$(Q1+Q2) and $P$(Q3+Q4). We find a significant relation between the $P$ distribution and the average values of all three indicators $S$, $T$ and $U$ with probabilities p<0.010 in all cases.[12] So we conclude that the Prognos Future Index ranking correlates well with each of our socioeconomic and population indicators.

## 5. Concluding Remarks

The study of the role of universities in the prosperity of cities and regions encounters two major problems. First, what is a reliable measurement of the diverse elements of prosperity; and second, given the wide variety of types of universities, what are the characteristics, particularly research performance of a university that really matter? In this study we focus on the research question: Is there a significant relation between having a university and a city's socioeconomic strength, its growth of the gross urban product, and its population size? And if so, what are the determining indicators of a university? In order to investigate this, we compiled a large database of city and university data: gross urban product and population data of nearly 200 German cities and 400 districts for the period 1997-2017. Data for the universities are derived from the CWTS bibliometric data system and supplemented with data on the number of students 1995-2020. Performance characteristics of universities are derived from the Leiden Ranking 2020. The socioeconomic strength of a city is determined with the urban scaling methodology.

Our study shows a significant relation between the *presence of a university in a city* and its socioeconomic indicators, particularly for larger cities. We find that for all three city indicators (socioeconomic strength, socioeconomic growth, and population growth) university cities are predominantly in the better half of the distribution function of these indicators.

In order to find *which university indicators* do have a significant relation with city indicators we developed two complementary data-analytical methods. In the first method the city indicators are leading and the analysis is focused on the question whether the university cities that are in the 'top' of a specific city indicator also are the cities that have a significantly higher score for one or more university indicators. In the second method the university indicators are leading, here the focus is on the question whether the cities of which the universities are 'top' in a specific university indicator also are the cities with a significantly higher values for one or more the city indicators.

We find that universities in cities with an above-average population growth are in general universities with a higher performance in scientific output (publication-based indicators), in scientific impact (citation-based indicators) and in scientific collaboration. Particularly collaboration with 'local' companies, i.e., companies in these cities and in their urban region relate to population growth. We also find indications that the size of the staff, which can be regarded as a pool of innovative people, could be a significant parameter in relation to the population growth of the city. For the socioeconomic strength of a city we find a relation particularly with the fractionally counted top impact indicators, both in absolute as well as in relative terms. These fractionally counted relative top impact indicators are a strong indicator of scientific quality. We conclude that university cities with an above-average socioeconomic strength have a higher probability that their university is a top-university as compared to cities in the below-average socioeconomic strength. Socioeconomic growth and city population growth appear to correlate significantly. This is to be expected, given the urban scaling relation between city population and the gross urban product of a city. Moreover, we find for socioeconomic growth of university cities a significant relation for relative number of scientific collaboration papers. We do not find a significant correlation between number of students and city indicators. This does not mean that size of student population does not matter: our group of universities consists of mostly large universities with high numbers of students, the average number of students is around 30,000. Apparently within that order of magnitude, further differences in student population do not give significant correlations with city indicators.

---

[12] Data and calculations are available in https://osf.io/4ru96/.

An interesting additional socioeconomic city indicator is provided by the ranking of cities and Kreise in the Prognos Future Index. This ranking index correlates well with each of our socioeconomic and population indicators. We find that university cities improved their socioeconomic ranking positions in the 15 years period between 2004 and 2019 significantly more than the other cities. In conclusion we have found a positive relation between having a major university and being a city with a relatively strong socioeconomic position and that this is especially the case for universities with higher values of their output and impact indicators. But this is certainly not an iron law for each university city, nor we make a statement about causality. An indication of a possible causal relation is that most of the major universities are centuries old whereas our socioeconomic indicators relate to recent times. This temporal precedence could suggest that cities with a major university will have a higher probability than cities without a major university to develop into a socioeconomically strong position. Finally we note that next to high quality research particularly applied research (including medical research) and related technological developments will probably also play an important role in the socioeconomic position of cities. To this end, we are currently investigating the patenting activities in cities together with a focus on the question whether the university indicators based on specifically applied research may relate stronger to the socioeconomic city indicators than the same university based on all university research.

## *Acknowledgements*

The author thanks Jos Winnink for the calculation of the bibliometric indicators for the German universities that are not included in the Leiden Ranking and for preparation and first analyses of the relevant patent data. The author also acknowledges Robert Tijssen for the calculation of university collaboration indicators.

# Supplementary Information

*Table S1. German (kreisfreie) cities (university cities, as far as present in the Leiden Ranking, in bold) ranked by $S$, with their national residual ($\xi_n$) and regional residual ($\xi_r$).*

| | $\xi_n$ | $\xi_r$ | $S$ | | $\xi_n$ | $\xi_r$ | $S$ |
|---:|---:|---:|---:|---:|---:|---:|---:|
| Wolfsburg | 1.17 | 1.23 | 1.20 | Erfurt | -0.25 | 0.13 | -0.06 |
| Ingolstadt | 0.99 | 0.81 | 0.90 | Remscheid | -0.33 | 0.19 | -0.07 |
| Schweinfurt | 0.75 | 0.61 | 0.68 | **Kiel** | -0.09 | -0.06 | -0.07 |
| **Erlangen** | 0.65 | 0.48 | 0.56 | **Oldenburg** | -0.10 | -0.06 | -0.08 |
| Coburg | 0.62 | 0.49 | 0.56 | **Dresden** | -0.28 | 0.09 | -0.09 |
| Ludwigshafen | 0.54 | 0.45 | 0.49 | Flensburg | -0.13 | -0.06 | -0.09 |
| **Regensburg** | 0.57 | 0.39 | 0.48 | **Freiburg** | 0.00 | -0.20 | -0.10 |
| **Bonn** | 0.34 | 0.52 | 0.43 | Pirmasens | -0.24 | 0.03 | -0.11 |
| **Düsseldorf** | 0.43 | 0.42 | 0.42 | Frankfurt (Oder) | -0.30 | 0.08 | -0.11 |
| Koblenz | 0.38 | 0.38 | 0.38 | Trier | -0.12 | -0.10 | -0.11 |
| Emden | 0.32 | 0.43 | 0.38 | Krefeld | -0.27 | 0.03 | -0.12 |
| **Darmstadt** | 0.41 | 0.34 | 0.37 | **Rostock** | -0.32 | 0.05 | -0.13 |
| **Stuttgart** | 0.49 | 0.25 | 0.37 | Augsburg | -0.03 | -0.24 | -0.13 |
| **Frankfurt** | 0.59 | 0.11 | 0.35 | Wilhelmshaven | -0.18 | -0.09 | -0.13 |
| Aschaffenburg | 0.42 | 0.27 | 0.35 | Neumünster | -0.20 | -0.12 | -0.16 |
| **Ulm** | 0.43 | 0.26 | 0.34 | Cottbus | -0.35 | 0.02 | -0.16 |
| Passau | 0.32 | 0.18 | 0.25 | **Berlin** | -0.36 | 0.02 | -0.17 |
| **München** | 0.37 | 0.10 | 0.24 | **Leipzig** | -0.36 | 0.01 | -0.17 |
| **Münster** | 0.11 | 0.31 | 0.21 | **Bielefeld** | -0.26 | -0.09 | -0.18 |
| **Bayreuth** | 0.28 | 0.12 | 0.20 | Mülheim a.d. Ruhr | -0.37 | 0.01 | -0.18 |
| Speyer | 0.07 | 0.29 | 0.18 | Hof | -0.12 | -0.26 | -0.19 |
| Memmingen | 0.23 | 0.10 | 0.17 | **Chemnitz** | -0.38 | 0.00 | -0.19 |
| Leverkusen | -0.04 | 0.36 | 0.16 | **Essen** | -0.19 | -0.19 | -0.19 |
| Bamberg | 0.24 | 0.08 | 0.16 | **Magdeburg** | -0.39 | -0.02 | -0.21 |
| Zweibrücken | -0.01 | 0.31 | 0.15 | Hagen | -0.39 | -0.04 | -0.21 |
| Mannheim | 0.25 | 0.04 | 0.14 | **Lübeck** | -0.24 | -0.20 | -0.22 |
| **Hamburg** | 0.18 | 0.10 | 0.14 | Worms | -0.27 | -0.18 | -0.22 |
| **Würzburg** | 0.22 | 0.05 | 0.14 | Wuppertal | -0.31 | -0.16 | -0.24 |
| Ansbach | 0.20 | 0.07 | 0.13 | Frankenthal (Pfalz) | -0.36 | -0.13 | -0.24 |
| **Karlsruhe** | 0.22 | 0.01 | 0.12 | Suhl | -0.43 | -0.06 | -0.25 |
| Wiesbaden | 0.23 | 0.00 | 0.11 | Solingen | -0.45 | -0.05 | -0.25 |
| Kempten (Allgäu) | 0.17 | 0.02 | 0.09 | Pforzheim | -0.17 | -0.34 | -0.25 |
| Landshut | 0.17 | 0.01 | 0.09 | Weimar | -0.45 | -0.08 | -0.26 |
| Weiden i.d.OPf. | 0.15 | 0.02 | 0.08 | Brandenburg | -0.45 | -0.08 | -0.27 |
| Amberg | 0.14 | 0.01 | 0.07 | Offenbach am Main | -0.27 | -0.28 | -0.28 |
| Salzgitter | 0.03 | 0.10 | 0.07 | Mönchengladbach | -0.41 | -0.16 | -0.28 |

| | | | | | | | | |
|---|---|---|---|---|---|---|---|---|
| Eisenach | -0.14 | 0.23 | 0.05 | | Bremerhaven | -0.32 | -0.26 | -0.29 |
| **Köln** | 0.14 | -0.05 | 0.05 | | Dessau-Roßlau | -0.49 | -0.12 | -0.30 |
| **Mainz** | 0.11 | -0.04 | 0.04 | | Kaufbeuren | -0.25 | -0.38 | -0.31 |
| Osnabrück | 0.01 | 0.06 | 0.03 | | **Dortmund** | -0.33 | -0.33 | -0.33 |
| Baden-Baden | 0.10 | -0.04 | 0.03 | | Neustadt a.d.Weinstr. | -0.44 | -0.24 | -0.34 |
| **Kaiserslautern** | -0.01 | 0.03 | 0.01 | | **Halle** | -0.53 | -0.16 | -0.35 |
| Straubing | 0.08 | -0.06 | 0.01 | | Schwabach | -0.29 | -0.42 | -0.36 |
| **Jena** | -0.18 | 0.19 | 0.01 | | **Duisburg** | -0.39 | -0.33 | -0.36 |
| **Nürnberg** | 0.11 | -0.12 | 0.00 | | Gera | -0.55 | -0.18 | -0.37 |
| **Heidelberg** | 0.09 | -0.10 | -0.01 | | **Bochum** | -0.45 | -0.30 | -0.37 |
| Heilbronn | 0.08 | -0.09 | -0.01 | | Hamm | -0.56 | -0.19 | -0.38 |
| Landau i.d. Pfalz | -0.13 | 0.11 | -0.01 | | Gelsenkirchen | -0.52 | -0.27 | -0.40 |
| **Potsdam** | -0.20 | 0.18 | -0.01 | | Fürth | -0.34 | -0.52 | -0.43 |
| **Bremen** | -0.01 | -0.03 | -0.02 | | Oberhausen | -0.59 | -0.27 | -0.43 |
| Rosenheim | 0.05 | -0.10 | -0.02 | | Herne | -0.75 | -0.34 | -0.54 |
| **Braunschweig** | -0.06 | -0.03 | -0.04 | | Bottrop | -0.82 | -0.32 | -0.57 |
| **Kassel** | 0.02 | -0.11 | -0.05 | | Delmenhorst | -0.75 | -0.66 | -0.70 |
| Schwerin | -0.24 | 0.13 | -0.06 | | | | | |

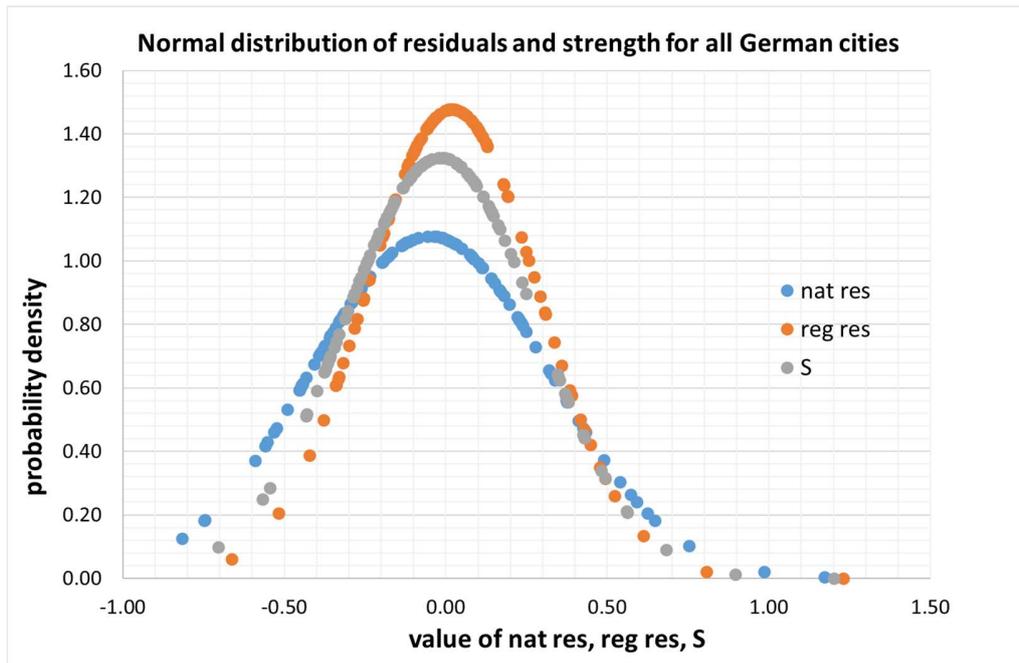

*Fig. S1 Normal distribution of the national residuals $\xi_n$ (nat res), the regional residuals $\xi_r$ (reg res) and of the strengths $S$ of all German (kreisfreie) cities.*

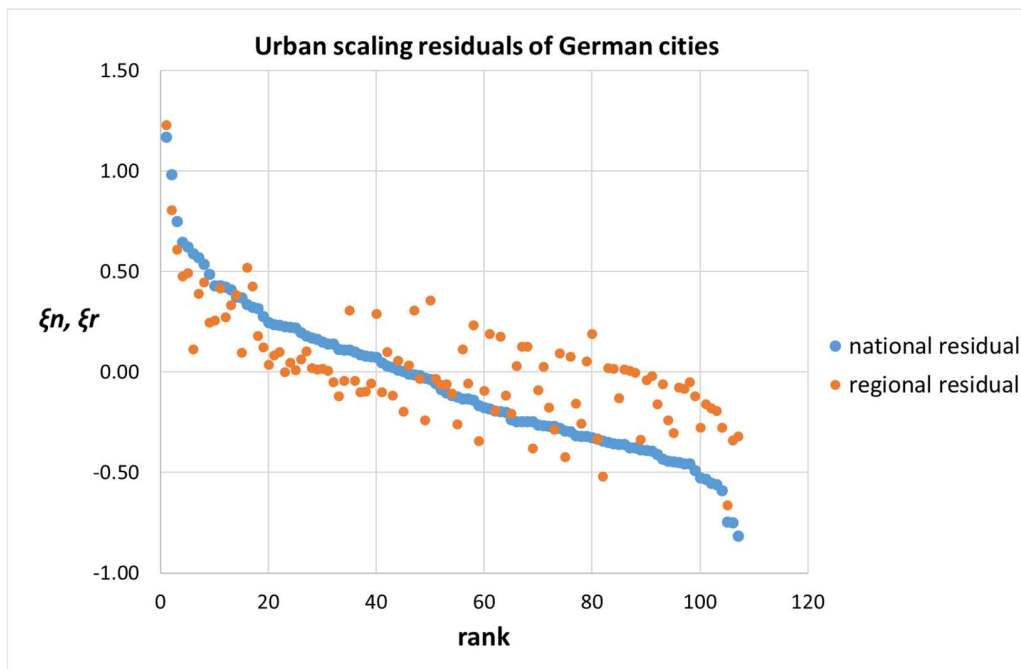

*Fig. S2 Ranking of the German (kreisfreie) cities according to their national urban scaling residuals ($\xi_n$) and also the regional residuals ($\xi_r$) for these cities are indicated (data in Table S1).*

*Table S2. German Kreise (with university cities, as far as present in the Leiden Ranking, in bold) ranked by $S$, with their national residual ($\xi_n$) and regional residual ($\xi_r$).*

| | $\xi_n$ | $\xi_r$ | $S$ | *cont'd* | $\xi_n$ | $\xi_r$ | $S$ |
|---|---|---|---|---|---|---|---|
| München, Kreis | 1.17 | 0.98 | 1.07 | Neuwied, Kreis | 0.01 | -0.04 | -0.01 |
| Dingolfing-Landau, Kreis | 0.90 | 0.74 | 0.82 | Lüchow-Dannenberg, Kreis | -0.05 | 0.02 | -0.01 |
| Böblingen, Kreis | 0.69 | 0.50 | 0.60 | Eichsfeld, Kreis | -0.10 | 0.07 | -0.01 |
| Main-Taunus-Kreis | 0.51 | 0.42 | 0.46 | Celle, Kreis | -0.04 | 0.01 | -0.01 |
| Altötting, Kreis | 0.53 | 0.37 | 0.45 | Nordsachsen, Kreis | -0.14 | 0.11 | -0.02 |
| Donau-Ries, Kreis | 0.52 | 0.36 | 0.44 | Uelzen, Kreis | -0.05 | 0.01 | -0.02 |
| Hohenlohekreis | 0.52 | 0.36 | 0.44 | Friesland, Kreis | -0.05 | 0.01 | -0.02 |
| Vechta, Kreis | 0.40 | 0.45 | 0.42 | **Tübingen, Kreis** | 0.07 | -0.11 | -0.02 |
| Biberach, Kreis | 0.50 | 0.33 | 0.42 | Goslar, Kreis | -0.05 | 0.01 | -0.02 |
| Tuttlingen, Kreis | 0.47 | 0.31 | 0.39 | Erding, Kreis | 0.06 | -0.10 | -0.02 |
| Bodenseekreis | 0.46 | 0.29 | 0.38 | Karlsruhe, Kreis | 0.07 | -0.12 | -0.02 |
| Gütersloh, Kreis | 0.34 | 0.37 | 0.36 | Wittmund, Kreis | -0.06 | 0.01 | -0.03 |
| Freising, Kreis | 0.43 | 0.26 | 0.34 | Mayen-Koblenz, Kreis | 0.01 | -0.07 | -0.03 |
| Hochtaunuskreis | 0.36 | 0.27 | 0.31 | Warendorf, Kreis | -0.04 | -0.02 | -0.03 |
| Günzburg, Kreis | 0.39 | 0.23 | 0.31 | Lörrach, Kreis | 0.06 | -0.12 | -0.03 |
| Heilbronn, Kreis | 0.41 | 0.21 | 0.31 | **Konstanz, Kreis** | 0.06 | -0.13 | -0.03 |
| Starnberg, Kreis | 0.39 | 0.22 | 0.30 | Neckar-Odenwald-Kreis | 0.05 | -0.12 | -0.03 |
| Rottweil, Kreis | 0.38 | 0.21 | 0.30 | Freyung-Grafenau, Kreis | 0.04 | -0.11 | -0.04 |
| Pfaffenhofen a.d.Ilm, Kreis | 0.36 | 0.20 | 0.28 | Wittenberg, Kreis | -0.13 | 0.06 | -0.04 |
| Teltow-Fläming, Kreis | 0.16 | 0.39 | 0.28 | Garmisch-Partenk., Kreis | 0.04 | -0.11 | -0.04 |
| Germersheim, Kreis | 0.28 | 0.27 | 0.27 | Schwalm-Eder-Kreis | -0.01 | -0.07 | -0.04 |
| Olpe, Kreis | 0.29 | 0.26 | 0.27 | Oberallgäu, Kreis | 0.04 | -0.12 | -0.04 |
| Groß-Gerau, Kreis | 0.33 | 0.22 | 0.27 | Nürnberger Land, Kreis | 0.04 | -0.12 | -0.04 |
| Saarpfalz-Kreis | 0.28 | 0.26 | 0.27 | Oberhavel, Kreis | -0.17 | 0.09 | -0.04 |
| Wesermarsch, Kreis | 0.23 | 0.29 | 0.26 | Weimarer Land, Kreis | -0.11 | 0.03 | -0.04 |
| Emsland, Kreis | 0.22 | 0.25 | 0.24 | Odenwaldkreis | -0.06 | -0.02 | -0.04 |
| **Saarbrücken, Reg.verband** | 0.30 | 0.17 | 0.23 | Aschaffenburg, Kreis | 0.04 | -0.13 | -0.05 |
| Rhön-Grabfeld, Kreis | 0.31 | 0.16 | 0.23 | Jerichower Land, Kreis | -0.12 | 0.03 | -0.05 |
| Vulkaneifel, Kreis | 0.18 | 0.28 | 0.23 | St. Wendel, Kreis | -0.07 | -0.02 | -0.05 |
| Minden-Lübbecke, Kreis | 0.22 | 0.24 | 0.23 | Hildburghausen, Kreis | -0.11 | 0.00 | -0.05 |
| Lindau (Bodensee), Kreis | 0.29 | 0.14 | 0.22 | Steinfurt, Kreis | -0.08 | -0.03 | -0.05 |
| Siegen-Wittgenstein, Kreis | 0.21 | 0.23 | 0.22 | **Mittelsachsen, Kreis** | -0.20 | 0.10 | -0.05 |
| Deggendorf, Kreis | 0.30 | 0.14 | 0.22 | Limburg-Weilburg, Kreis | -0.03 | -0.08 | -0.06 |
| Rastatt, Kreis | 0.31 | 0.13 | 0.22 | Emmendingen, Kreis | 0.03 | -0.14 | -0.06 |
| Rhein-Kreis Neuss, Kreis | 0.19 | 0.24 | 0.21 | Sömmerda, Kreis | -0.12 | 0.00 | -0.06 |
| Rhein-Hunsrück-Kreis | 0.20 | 0.22 | 0.21 | Ennepe-Ruhr-Kreis | -0.07 | -0.04 | -0.06 |
| Hersfeld-Rotenburg, Kreis | 0.20 | 0.21 | 0.21 | Unstrut-Hainich-Kreis | -0.15 | 0.02 | -0.06 |
| Freudenstadt, Kreis | 0.28 | 0.12 | 0.20 | Mecklenb. Seenplatte, Kreis | -0.20 | 0.08 | -0.06 |
| Dahme-Spreewald, Kreis | 0.09 | 0.31 | 0.20 | Kassel, Kreis | -0.02 | -0.11 | -0.07 |
| Weilheim-Schongau, Kreis | 0.28 | 0.11 | 0.19 | Waldshut, Kreis | 0.02 | -0.15 | -0.07 |
| Saalekreis | 0.07 | 0.31 | 0.19 | Vogelsbergkreis | -0.08 | -0.06 | -0.07 |
| Offenbach, Kreis | 0.26 | 0.12 | 0.19 | Stendal, Kreis | -0.16 | 0.02 | -0.07 |
| Spree-Neiße, Kreis | 0.09 | 0.28 | 0.19 | Leer, Kreis | -0.10 | -0.05 | -0.07 |
| Schwäbisch Hall, Kreis | 0.27 | 0.10 | 0.18 | Oder-Spree, Kreis | -0.19 | 0.04 | -0.07 |
| Lichtenfels, Kreis | 0.25 | 0.11 | 0.18 | Görlitz, Kreis | -0.22 | 0.06 | -0.08 |
| Ravensburg, Kreis | 0.27 | 0.09 | 0.18 | Merzig-Wadern, Kreis | -0.09 | -0.07 | -0.08 |
| Neumarkt i.d.OPf., Kreis | 0.26 | 0.10 | 0.18 | Elbe-Elster, Kreis | -0.17 | 0.00 | -0.08 |
| Traunstein, Kreis | 0.26 | 0.09 | 0.18 | Bad Tölz-Wolfratsh., Kreis | -0.01 | -0.17 | -0.09 |
| Mettmann, Kreis | 0.14 | 0.20 | 0.17 | Altenkirchen (W.), Kreis | -0.08 | -0.09 | -0.09 |
| Cochem-Zell, Kreis | 0.12 | 0.22 | 0.17 | Osnabrück, Kreis | -0.10 | -0.07 | -0.09 |
| Nordfriesland, Kreis | 0.14 | 0.19 | 0.17 | Neustadt a.d.A.-B. W., Kreis | -0.01 | -0.16 | -0.09 |
| Ludwigsburg, Kreis | 0.26 | 0.06 | 0.16 | Lüneburg, Kreis | -0.11 | -0.06 | -0.09 |
| Waldeck-Frankenberg, Kreis | 0.18 | 0.14 | 0.16 | SalzKreis | -0.21 | 0.03 | -0.09 |
| Ostalbkreis | 0.25 | 0.06 | 0.15 | Rems-Murr-Kreis | 0.01 | -0.19 | -0.09 |
| Holzminden, Kreis | 0.12 | 0.18 | 0.15 | Main-Kinzig-Kreis | 0.00 | -0.18 | -0.09 |
| Tirschenreuth, Kreis | 0.22 | 0.07 | 0.15 | Höxter, Kreis | -0.08 | -0.11 | -0.09 |
| **Region Hannover, Kreis** | 0.14 | 0.15 | 0.15 | Lippe, Kreis | -0.11 | -0.07 | -0.09 |
| Kulmbach, Kreis | 0.21 | 0.07 | 0.14 | Altmarkkreis Salzwedel | -0.17 | -0.02 | -0.09 |
| Hameln-Pyrmont, Kreis | 0.12 | 0.17 | 0.14 | Leipzig, Kreis | -0.23 | 0.05 | -0.09 |
| Bernkastel-Wittlich, Kreis | 0.13 | 0.15 | 0.14 | Unna, Kreis | -0.12 | -0.07 | -0.09 |
| Main-Tauber-Kreis | 0.22 | 0.06 | 0.14 | Kleve, Kreis | -0.11 | -0.08 | -0.09 |
| Kitzingen, Kreis | 0.21 | 0.06 | 0.14 | Südliche Weinstraße, Kreis | -0.11 | -0.09 | -0.10 |
| Neuburg-Schrobenh., Kreis | 0.21 | 0.06 | 0.13 | Rendsburg-Eckernf., Kreis | -0.12 | -0.08 | -0.10 |
| Fulda, Kreis | 0.17 | 0.09 | 0.13 | Rosenheim, Kreis | -0.01 | -0.19 | -0.10 |
| Unterallgäu, Kreis | 0.21 | 0.05 | 0.13 | Bautzen, Kreis | -0.25 | 0.05 | -0.10 |
| Cloppenburg, Kreis | 0.10 | 0.15 | 0.13 | Alb-Donau-Kreis | -0.02 | -0.19 | -0.11 |
| Schwarzwald-Baar-Kreis | 0.21 | 0.04 | 0.12 | Ansbach, Kreis | -0.02 | -0.19 | -0.11 |
| Reutlingen, Kreis | 0.21 | 0.03 | 0.12 | Potsdam-Mittelmark, Kreis | -0.23 | 0.02 | -0.11 |
| Uckermark, Kreis | 0.03 | 0.22 | 0.12 | BurgenKreis | -0.23 | 0.01 | -0.11 |
| Neu-Ulm, Kreis | 0.21 | 0.04 | 0.12 | Werra-Meißner-Kreis | -0.12 | -0.10 | -0.11 |
| Sigmaringen, Kreis | 0.20 | 0.04 | 0.12 | Düren, Kreis | -0.12 | -0.10 | -0.11 |
| Donnersbergkreis | 0.07 | 0.14 | 0.11 | Coesfeld, Kreis | -0.11 | -0.11 | -0.11 |
| Miesbach, Kreis | 0.18 | 0.03 | 0.11 | Pinneberg, Kreis | -0.13 | -0.10 | -0.11 |
| Hof, Kreis | 0.18 | 0.03 | 0.11 | Coburg, Kreis | -0.04 | -0.19 | -0.11 |
| Oberbergischer Kreis | 0.10 | 0.12 | 0.11 | VogtKreis | -0.25 | 0.02 | -0.12 |
| HochsauerKreis | 0.10 | 0.11 | 0.11 | Forchheim, Kreis | -0.04 | -0.19 | -0.12 |
| Cham, Kreis | 0.19 | 0.03 | 0.11 | Roth, Kreis | -0.04 | -0.20 | -0.12 |
| Dillingen a.d.Donau, Kreis | 0.18 | 0.03 | 0.10 | Enzkreis | -0.03 | -0.21 | -0.12 |
| Rotenburg (Wümme), Kreis | 0.08 | 0.13 | 0.10 | Bergstraße, Kreis | -0.06 | -0.17 | -0.12 |
| Märkischer Kreis | 0.08 | 0.13 | 0.10 | Viersen, Kreis | -0.13 | -0.11 | -0.12 |
| Steinburg, Kreis | 0.08 | 0.13 | 0.10 | Rhein-Lahn-Kreis | -0.12 | -0.13 | -0.12 |
| Sonneberg, Kreis | 0.05 | 0.15 | 0.10 | Straubing-Bogen, Kreis | -0.05 | -0.20 | -0.13 |
| Wunsiedel Fichtelgeb., Kreis | 0.17 | 0.03 | 0.10 | Hildesheim, Kreis | -0.15 | -0.11 | -0.13 |
| Heidekreis, Kreis | 0.08 | 0.13 | 0.10 | Aurich, Kreis | -0.15 | -0.11 | -0.13 |
| Schwandorf, Kreis | 0.18 | 0.02 | 0.10 | Saale-Holzland-Kreis | -0.20 | -0.06 | -0.13 |
| Erlangen-Höchstadt, Kreis | 0.18 | 0.02 | 0.10 | Kreis Rostock | -0.26 | 0.00 | -0.13 |
| Main-Spessart, Kreis | 0.18 | 0.02 | 0.10 | Ahrweiler, Kreis | -0.13 | -0.14 | -0.13 |
| Mainz-Bingen, Kreis | 0.14 | 0.06 | 0.10 | Passau, Kreis | -0.05 | -0.22 | -0.14 |
| **Paderborn, Kreis** | 0.09 | 0.11 | 0.10 | Schleswig-Flensburg, Kreis | -0.16 | -0.12 | -0.14 |
| Borken, Kreis | 0.08 | 0.12 | 0.10 | Calw, Kreis | -0.06 | -0.23 | -0.14 |
| Dithmarschen, Kreis | 0.07 | 0.12 | 0.10 | Rhein-Neckar-Kreis | -0.04 | -0.25 | -0.15 |
| Birkenfeld, Kreis | 0.07 | 0.12 | 0.10 | Dachau, Kreis | -0.06 | -0.23 | -0.15 |
| Heidenheim, Kreis | 0.18 | 0.01 | 0.10 | **Vorpomm.-Greifswald, Kreis** | -0.28 | -0.01 | -0.15 |
| **Städteregion Aachen, Kreis** | 0.06 | 0.13 | 0.09 | Greiz, Kreis | -0.23 | -0.07 | -0.15 |
| Westerwaldkreis | 0.13 | 0.06 | 0.09 | Euskirchen, Kreis | -0.15 | -0.16 | -0.15 |
| Ilm-Kreis | 0.00 | 0.18 | 0.09 | Würzburg, Kreis | -0.07 | -0.24 | -0.16 |
| Saale-Orla-Kreis | 0.02 | 0.16 | 0.09 | Vorpommern-Rügen, Kreis | -0.29 | -0.03 | -0.16 |
| Grafschaft Bentheim, Kreis | 0.06 | 0.11 | 0.09 | Alzey-Worms, Kreis | -0.16 | -0.16 | -0.16 |
| Zwickau, Kreis | -0.07 | 0.23 | 0.08 | Amberg-Sulzbach, Kreis | -0.08 | -0.24 | -0.16 |
| Kronach, Kreis | 0.15 | 0.01 | 0.08 | Neunkirchen, Kreis | -0.16 | -0.17 | -0.16 |
| **Marburg-Biedenkopf, Kreis** | 0.13 | 0.03 | 0.08 | Schaumburg, Kreis | -0.19 | -0.14 | -0.16 |
| Nienburg (Weser), Kreis | 0.05 | 0.11 | 0.08 | Augsburg, Kreis | -0.08 | -0.26 | -0.17 |
| Ostallgäu, Kreis | 0.16 | -0.01 | 0.07 | Ludwigslust-Parchim, Kreis | -0.30 | -0.04 | -0.17 |
| Soest, Kreis | 0.06 | 0.08 | 0.07 | Harz, Kreis | -0.30 | -0.04 | -0.17 |
| Stormarn, Kreis | 0.05 | 0.09 | 0.07 | Breisgau-Hochschw., Kreis | -0.08 | -0.26 | -0.17 |
| Rhein-Erft-Kreis | 0.04 | 0.10 | 0.07 | Wesel, Kreis | -0.20 | -0.14 | -0.17 |
| Verden, Kreis | 0.04 | 0.09 | 0.07 | Ostholstein, Kreis | -0.20 | -0.15 | -0.17 |
| Ortenaukreis | 0.16 | -0.03 | 0.06 | Nordwestmecklenburg, Kreis | -0.29 | -0.07 | -0.18 |
| **Göttingen, Kreis** | 0.05 | 0.08 | 0.06 | Altenburger Land, Kreis | -0.25 | -0.10 | -0.18 |
| Saarlouis, Kreis | 0.10 | 0.03 | 0.06 | Aichach-Friedberg, Kreis | -0.10 | -0.26 | -0.18 |
| Esslingen, Kreis | 0.16 | -0.04 | 0.06 | Wetteraukreis | -0.12 | -0.24 | -0.18 |
| Herford, Kreis | 0.05 | 0.06 | 0.06 | Rhein-Sieg-Kreis | -0.22 | -0.15 | -0.18 |
| Miltenberg, Kreis | 0.14 | -0.02 | 0.06 | Kyffhäuserkreis | -0.25 | -0.12 | -0.19 |
| Börde, Kreis | -0.07 | 0.16 | 0.05 | Oldenburg, Kreis | -0.22 | -0.17 | -0.19 |
| Mühldorf a.Inn, Kreis | 0.13 | -0.03 | 0.05 | Sächsische Schweiz-O., Kreis | -0.34 | -0.07 | -0.21 |
| Bad Kissingen, Kreis | 0.12 | -0.03 | 0.05 | Darmstadt-Dieburg, Kreis | -0.15 | -0.27 | -0.21 |
| Regen, Kreis | 0.12 | -0.03 | 0.04 | Rheinisch-Bergischer Kreis | -0.23 | -0.21 | -0.22 |
| Landshut, Kreis | 0.12 | -0.04 | 0.04 | Barnim, Kreis | -0.33 | -0.10 | -0.22 |
| Berchtesgadener Land, Kreis | 0.11 | -0.04 | 0.04 | Erzgebirgskreis | -0.38 | -0.06 | -0.22 |
| Landsberg am Lech, Kreis | 0.12 | -0.04 | 0.04 | Bamberg, Kreis | -0.15 | -0.32 | -0.24 |
| Anhalt-Bitterfeld, Kreis | -0.08 | 0.14 | 0.03 | Heinsberg, Kreis | -0.25 | -0.24 | -0.24 |
| Eichstätt, Kreis | 0.11 | -0.05 | 0.03 | Rheingau-Taunus-Kreis | -0.22 | -0.27 | -0.24 |
| **Giessen, Kreis** | 0.08 | -0.03 | 0.03 | Harburg, Kreis | -0.27 | -0.23 | -0.25 |
| Gotha, Kreis | -0.07 | 0.13 | 0.03 | Kaiserslautern, Kreis | -0.27 | -0.25 | -0.26 |
| Ammerland, Kreis | 0.00 | 0.05 | 0.03 | Bad Dürkheim, Kreis | -0.26 | -0.27 | -0.26 |
| Schmalkalden-Mein., Kreis | -0.07 | 0.12 | 0.02 | Mansfeld-Südharz, Kreis | -0.38 | -0.17 | -0.28 |
| Zollernalbkreis | 0.11 | -0.06 | 0.02 | Märkisch-Oderland, Kreis | -0.40 | -0.16 | -0.28 |
| Haßberge, Kreis | 0.09 | -0.06 | 0.02 | Gifhorn, Kreis | -0.30 | -0.26 | -0.28 |
| Stade, Kreis | -0.01 | 0.04 | 0.02 | Peine, Kreis | -0.31 | -0.26 | -0.28 |
| Prignitz, Kreis | -0.05 | 0.08 | 0.01 | Fürstenfeldbruck, Kreis | -0.20 | -0.38 | -0.29 |
| Wartburgkreis | -0.08 | 0.11 | 0.01 | Havelland, Kreis | -0.40 | -0.18 | -0.29 |
| Kelheim, Kreis | 0.09 | -0.07 | 0.01 | Helmstedt, Kreis | -0.33 | -0.27 | -0.30 |
| Weißenburg-Gunzenh., Kreis | 0.08 | -0.07 | 0.01 | Herzogtum Lauenburg, Kreis | -0.32 | -0.28 | -0.30 |
| Oberspreewald-Lausitz, Kreis | -0.08 | 0.10 | 0.01 | Cuxhaven, Kreis | -0.32 | -0.28 | -0.30 |
| Eifelkreis Bitburg-Prüm | -0.01 | 0.02 | 0.01 | Recklinghausen, Kreis | -0.34 | -0.27 | -0.30 |
| Saalfeld-Rudolstadt, Kreis | -0.08 | 0.09 | 0.00 | Regensburg, Kreis | -0.22 | -0.39 | -0.31 |
| Diepholz, Kreis | -0.02 | 0.02 | 0.00 | Fürth, Kreis | -0.23 | -0.39 | -0.31 |
| Bad Kreuznach, Kreis | 0.02 | -0.02 | 0.00 | Wolfenbüttel, Kreis | -0.36 | -0.31 | -0.33 |
| Meißen, Kreis | -0.14 | 0.13 | 0.00 | Osterholz, Kreis | -0.37 | -0.31 | -0.34 |
| Neustadt a.d.Waldn., Kreis | 0.07 | -0.08 | 0.00 | Plön, Kreis | -0.38 | -0.33 | -0.35 |
| Northeim, Kreis | -0.03 | 0.02 | 0.00 | Schweinfurt, Kreis | -0.29 | -0.45 | -0.37 |
| Segeberg, Kreis | -0.02 | 0.02 | 0.00 | Bayreuth, Kreis | -0.31 | -0.46 | -0.39 |
| Nordhausen, Kreis | -0.08 | 0.07 | -0.01 | Trier-Saarburg, Kreis | -0.38 | -0.41 | -0.40 |
| Lahn-Dill-Kreis | 0.04 | -0.06 | -0.01 | Kusel, Kreis | -0.46 | -0.38 | -0.42 |
| Rottal-Inn, Kreis | 0.07 | -0.09 | -0.01 | Rhein-Pfalz-Kreis | -0.48 | -0.51 | -0.50 |
| Ebersberg, Kreis | 0.07 | -0.09 | -0.01 | Südwestpfalz, Kreis | -0.55 | -0.52 | -0.53 |
| Göppingen, Kreis | 0.08 | -0.10 | -0.01 | | | | |
| Ostprignitz-Ruppin, Kreis | -0.09 | 0.07 | -0.01 | | | | |

Table S3. Number of inhabitants ($N$, year 2019), socioeconomic strength $S$, growth of the gross urban product $T$, and population growth $U$ for all cities between 50,000 and 100,000 inhabitants (first table) and cities above 100,00 (second table). University cities, as far as present in the Leiden Ranking, in bold). Cities are ranked by S.

| 110 cities 50,000-100,000 | $N$ | $S$ | $T$ | $U$ |
|---|---|---|---|---|
| Schweinfurt | 53,426 | 0.68 | 1.90 | 0.98 |
| Böblingen | 50,161 | 0.60 | 2.01 | 1.09 |
| Sindelfingen | 64,905 | 0.60 | 2.01 | 1.07 |
| Friedrichshafen | 61,283 | 0.38 | 2.10 | 1.07 |
| Aschaffenburg | 71,002 | 0.35 | 1.88 | 1.05 |
| Bad Homburg vor der Höhe | 54,227 | 0.31 | 1.42 | 1.03 |
| Rüsselsheim am Main | 65,881 | 0.27 | 1.48 | 1.11 |
| Passau | 52,803 | 0.25 | 1.72 | 1.04 |
| Lingen (Ems) | 54,708 | 0.24 | 1.90 | 1.06 |
| Minden | 81,716 | 0.23 | 1.58 | 0.98 |
| Meerbusch | 56,415 | 0.21 | 1.75 | 1.02 |
| Dormagen | 64,340 | 0.21 | 1.75 | 1.02 |
| Grevenbroich | 63,743 | 0.21 | 1.75 | 0.99 |
| **Bayreuth** | 74,783 | 0.20 | 1.60 | 1.01 |
| Speyer | 50,561 | 0.18 | 1.83 | 1.02 |
| Ravensburg | 50,897 | 0.18 | 1.82 | 1.07 |
| Langenfeld (Rheinland) | 59,178 | 0.17 | 1.51 | 1.01 |
| Hilden | 55,625 | 0.17 | 1.51 | 0.99 |
| Ratingen | 87,520 | 0.17 | 1.51 | 0.96 |
| Velbert | 81,842 | 0.17 | 1.51 | 0.91 |
| Bamberg | 77,373 | 0.16 | 1.66 | 1.12 |
| Ludwigsburg | 93,584 | 0.16 | 1.88 | 1.07 |
| Langenhagen | 54,652 | 0.15 | 1.60 | 1.11 |
| Aalen | 68,393 | 0.15 | 1.91 | 1.04 |
| Schwäbisch Gmünd | 61,137 | 0.15 | 1.91 | 0.98 |
| Garbsen | 61,032 | 0.15 | 1.60 | 0.96 |
| Hameln | 57,434 | 0.14 | 1.44 | 0.98 |
| Fulda | 68,635 | 0.13 | 1.75 | 1.10 |
| Neu-Ulm | 58,978 | 0.12 | 1.87 | 1.18 |
| Villingen-Schwenningen | 85,707 | 0.12 | 1.69 | 1.06 |
| Gummersbach | 50,952 | 0.11 | 1.60 | 0.96 |
| Arnsberg | 73,456 | 0.11 | 1.49 | 0.95 |
| Bocholt | 71,113 | 0.10 | 1.84 | 0.99 |
| Iserlohn | 92,174 | 0.10 | 1.49 | 0.93 |
| Lüdenscheid | 72,313 | 0.10 | 1.49 | 0.90 |
| Menden (Sauerland) | 52,608 | 0.10 | 1.49 | 0.89 |
| Kempten (Allgäu) | 69,151 | 0.09 | 1.80 | 1.13 |
| Nordhorn | 53,711 | 0.09 | 1.84 | 1.03 |
| Eschweiler | 56,482 | 0.09 | 1.60 | 1.03 |
| Stolberg (Rheinland) | 56,466 | 0.09 | 1.60 | 0.96 |
| Landshut | 73,411 | 0.09 | 1.56 | 1.25 |
| Zwickau | 88,690 | 0.08 | 1.68 | 0.86 |
| **Marburg** | 77,129 | 0.08 | 1.57 | 1.00 |
| Hürth | 59,731 | 0.07 | 1.58 | 1.12 |
| Frechen | 52,439 | 0.07 | 1.58 | 1.12 |
| Kerpen | 66,702 | 0.07 | 1.58 | 1.06 |
| Pulheim | 54,194 | 0.07 | 1.58 | 1.02 |
| Lippstadt | 67,952 | 0.07 | 1.59 | 1.02 |
| Erftstadt | 50,010 | 0.07 | 1.58 | 0.99 |
| Bergheim | 61,601 | 0.07 | 1.58 | 0.97 |
| Offenburg | 59,967 | 0.06 | 1.69 | 1.05 |
| Esslingen am Neckar | 94,145 | 0.06 | 1.62 | 1.05 |
| Herford | 66,638 | 0.06 | 1.42 | 1.02 |
| Baden-Baden | 55,185 | 0.03 | 1.68 | 1.05 |
| **Gießen** | 89,802 | 0.03 | 1.44 | 1.23 |
| Bad Kreuznach | 51,170 | 0.00 | 1.59 | 1.19 |
| Norderstedt | 79,357 | 0.00 | 1.46 | 1.11 |
| Göppingen | 57,813 | -0.01 | 1.51 | 1.01 |
| Wetzlar | 52,955 | -0.01 | 1.55 | 1.01 |
| Neuwied | 64,765 | -0.01 | 1.52 | 0.97 |
| Celle | 69,540 | -0.01 | 1.47 | 0.96 |
| **Tübingen** | 91,506 | -0.02 | 1.90 | 1.13 |
| Goslar | 50,554 | -0.02 | 1.40 | 1.14 |
| Rosenheim | 63,551 | -0.02 | 1.41 | 1.08 |
| Ahlen | 52,503 | -0.03 | 1.54 | 0.94 |
| **Konstanz** | 84,911 | -0.03 | 1.78 | 1.09 |
| Ibbenbüren | 51,822 | -0.05 | 1.78 | 1.06 |
| Rheine | 76,218 | -0.05 | 1.78 | 1.00 |
| Schwerin Landeshauptstadt | 95,653 | -0.06 | 1.48 | 0.94 |
| Witten | 96,459 | -0.06 | 1.48 | 0.93 |
| Hattingen | 54,438 | -0.06 | 1.48 | 0.93 |
| Neubrandenburg | 63,761 | -0.06 | 1.42 | 0.87 |
| Görlitz | 55,980 | -0.08 | 1.48 | 0.91 |
| Lüneburg | 75,711 | -0.09 | 1.72 | 1.12 |
| Hanau | 96,492 | -0.09 | 1.63 | 1.09 |
| Waiblingen | 55,604 | -0.09 | 1.56 | 1.08 |
| Kleve | 52,388 | -0.09 | 1.81 | 1.08 |
| Detmold | 74,254 | -0.09 | 1.42 | 1.01 |
| Bad Salzuflen | 54,254 | -0.09 | 1.42 | 0.99 |

| City | N | S | T | U |
|---|---|---|---|---|
| Lünen | 86,348 | -0.09 | 1.69 | 0.94 |
| Unna | 58,936 | -0.09 | 1.69 | 0.83 |
| Flensburg | 90,164 | -0.09 | 1.32 | 1.07 |
| Frankfurt (Oder) | 57,751 | -0.11 | 1.19 | 0.80 |
| Düren | 91,216 | -0.11 | 1.46 | 0.99 |
| Willich | 50,391 | -0.12 | 1.50 | 1.00 |
| Viersen | 77,102 | -0.12 | 1.50 | 1.00 |
| Plauen | 64,597 | -0.12 | 1.45 | 0.90 |
| Wilhelmshaven | 76,089 | -0.13 | 1.36 | 0.89 |
| **Greifswald** | 59,232 | -0.15 | 1.60 | 1.09 |
| Euskirchen | 58,381 | -0.15 | 1.51 | 1.09 |
| Neumünster | 80,196 | -0.16 | 1.39 | 1.00 |
| Stralsund | 59,418 | -0.16 | 1.56 | 0.98 |
| Cottbus | 99,678 | -0.16 | 1.40 | 0.92 |
| Wesel | 60,230 | -0.17 | 1.66 | 0.97 |
| Dinslaken | 67,373 | -0.17 | 1.66 | 0.95 |
| Troisdorf | 74,953 | -0.18 | 1.69 | 1.03 |
| Sankt Augustin | 55,847 | -0.18 | 1.69 | 1.01 |
| Worms | 83,542 | -0.22 | 1.61 | 1.04 |
| Weimar | 65,228 | -0.26 | 1.52 | 1.04 |
| Brandenburg an der Havel | 72,184 | -0.27 | 1.49 | 0.93 |
| Gladbeck | 75,610 | -0.30 | 1.55 | 0.97 |
| Castrop-Rauxel | 73,343 | -0.30 | 1.55 | 0.93 |
| Herten | 61,821 | -0.30 | 1.55 | 0.92 |
| Dorsten | 74,704 | -0.30 | 1.55 | 0.92 |
| Marl | 84,067 | -0.30 | 1.55 | 0.90 |
| Dessau-Roßlau | 80,103 | -0.30 | 1.40 | 0.96 |
| Wolfenbüttel | 52,165 | -0.33 | 1.68 | 0.95 |
| Neustadt an der Weinstraße | 53,264 | -0.34 | 1.35 | 0.99 |
| Gera | 93,125 | -0.37 | 1.30 | 0.83 |
| Delmenhorst | 77,559 | -0.70 | 1.55 | 1.01 |

| *81 cities >100,000* | $N$ | $S$ | $T$ | $U$ |
|---|---|---|---|---|
| Wolfsburg | 124,371 | 1.20 | 2.15 | 1.02 |
| Ingolstadt | 137,392 | 0.90 | 3.80 | 1.19 |
| **Erlangen** | 112,528 | 0.56 | 2.02 | 1.12 |
| Ludwigshafen | 172,253 | 0.49 | 1.46 | 1.05 |
| **Regensburg** | 153,094 | 0.48 | 1.99 | 1.22 |
| **Bonn** | 329,673 | 0.43 | 1.42 | 1.09 |
| **Düsseldorf** | 621,877 | 0.42 | 1.52 | 1.09 |
| Koblenz | 114,052 | 0.38 | 1.42 | 1.06 |
| **Darmstadt** | 159,878 | 0.37 | 1.59 | 1.16 |

| City | Population | Col3 | Col4 | Col5 |
|---|---|---|---|---|
| **Stuttgart** | 635,911 | 0.37 | 1.60 | 1.09 |
| Gütersloh | 100,861 | 0.36 | 1.74 | 1.06 |
| **Frankfurt** | 763,380 | 0.35 | 1.50 | 1.18 |
| **Ulm** | 126,790 | 0.34 | 1.72 | 1.08 |
| **München** | 1,484,226 | 0.24 | 1.75 | 1.23 |
| **Saarbrücken** | 180,374 | 0.23 | 1.47 | 0.98 |
| Siegen | 102,770 | 0.22 | 1.52 | 0.95 |
| **Münster** | 315,293 | 0.21 | 1.53 | 1.19 |
| Neuss | 153,896 | 0.21 | 1.75 | 1.03 |
| Leverkusen | 163,729 | 0.16 | 1.18 | 1.02 |
| **Hannover** | 536,925 | 0.15 | 1.60 | 1.04 |
| Mannheim | 310,658 | 0.14 | 1.59 | 1.01 |
| **Hamburg** | 1,847,253 | 0.14 | 1.57 | 1.08 |
| **Würzburg** | 127,934 | 0.14 | 1.52 | 1.00 |
| Reutlingen | 115,865 | 0.12 | 1.71 | 1.05 |
| **Karlsruhe** | 312,060 | 0.12 | 1.54 | 1.12 |
| Wiesbaden | 278,474 | 0.11 | 1.61 | 1.03 |
| **Paderborn** | 151,633 | 0.10 | 1.76 | 1.09 |
| **Aachen** | 248,960 | 0.09 | 1.60 | 1.02 |
| Salzgitter | 104,291 | 0.07 | 1.58 | 0.93 |
| **Göttingen** | 118,911 | 0.06 | 1.51 | 0.96 |
| **Köln** | 1,087,863 | 0.05 | 1.57 | 1.13 |
| **Mainz** | 218,578 | 0.04 | 1.46 | 1.20 |
| Osnabrück | 165,251 | 0.03 | 1.45 | 1.01 |
| **Kaiserslautern** | 100,030 | 0.01 | 1.48 | 1.00 |
| **Jena** | 111,343 | 0.01 | 2.28 | 1.11 |
| **Nürnberg** | 518,370 | 0.00 | 1.65 | 1.06 |
| **Heidelberg** | 161,485 | -0.01 | 1.74 | 1.15 |
| Heilbronn | 126,592 | -0.01 | 1.41 | 1.06 |
| **Potsdam** | 180,334 | -0.01 | 1.89 | 1.39 |
| **Bremen** | 567,559 | -0.02 | 1.58 | 1.05 |
| **Braunschweig** | 249,406 | -0.04 | 1.64 | 1.01 |
| **Kassel** | 202,137 | -0.05 | 1.45 | 1.04 |
| Erfurt | 213,981 | -0.06 | 1.55 | 1.07 |
| Remscheid | 111,338 | -0.07 | 1.22 | 0.93 |
| **Kiel** | 246,794 | -0.07 | 1.47 | 1.06 |
| **Oldenburg** | 169,077 | -0.08 | 1.49 | 1.09 |
| **Dresden** | 556,780 | -0.09 | 1.87 | 1.17 |
| **Freiburg** | 231,195 | -0.10 | 1.74 | 1.13 |
| Trier | 111,528 | -0.11 | 1.49 | 1.12 |
| Krefeld | 227,417 | -0.12 | 1.27 | 0.95 |
| Hildesheim | 101,693 | -0.13 | 1.47 | 0.98 |
| **Rostock** | 209,191 | -0.13 | 1.65 | 1.04 |
| Augsburg | 296,582 | -0.13 | 1.51 | 1.16 |
| **Berlin** | 3,669,491 | -0.17 | 1.57 | 1.08 |
| Moers | 103,902 | -0.17 | 1.66 | 0.97 |
| **Leipzig** | 593,145 | -0.17 | 1.87 | 1.20 |
| **Bielefeld** | 334,195 | -0.18 | 1.56 | 1.04 |
| Mülheim | 170,632 | -0.18 | 1.34 | 0.99 |
| **Chemnitz** | 246,334 | -0.19 | 1.42 | 0.95 |
| **Essen** | 582,760 | -0.19 | 1.37 | 0.98 |
| **Magdeburg** | 237,565 | -0.21 | 1.48 | 1.03 |
| Hagen | 188,686 | -0.21 | 1.30 | 0.93 |
| **Lübeck** | 216,530 | -0.22 | 1.53 | 1.01 |
| Bergisch Gladbach | 111,846 | -0.22 | 1.49 | 1.06 |
| Wuppertal | 355,100 | -0.24 | 1.38 | 0.97 |
| Solingen | 159,245 | -0.25 | 1.58 | 0.97 |
| Pforzheim | 125,957 | -0.25 | 1.51 | 1.08 |
| Offenbach am Main | 130,280 | -0.28 | 1.30 | 1.11 |
| Mönchengladbach | 261,034 | -0.28 | 1.51 | 0.99 |
| Bremerhaven | 113,643 | -0.29 | 1.37 | 0.94 |
| Recklinghausen | 111,397 | -0.30 | 1.55 | 0.89 |
| **Dortmund** | 588,250 | -0.33 | 1.56 | 1.00 |
| **Halle** | 238,762 | -0.35 | 1.23 | 0.96 |
| **Duisburg** | 498,686 | -0.36 | 1.43 | 0.97 |
| **Bochum** | 365,587 | -0.37 | 1.28 | 0.93 |
| Hamm | 179,916 | -0.38 | 1.58 | 0.99 |
| Gelsenkirchen | 259,645 | -0.40 | 1.35 | 0.93 |
| Fürth | 128,497 | -0.43 | 1.55 | 1.16 |
| Oberhausen | 210,764 | -0.43 | 1.39 | 0.95 |
| Herne | 156,449 | -0.54 | 1.36 | 0.90 |
| Bottrop | 117,565 | -0.57 | 1.40 | 0.97 |

*Table S4. Number of university cities and other cities for each quartile of the socioeconomic growth (T) distribution and of the population growth (U) distribution (Left side: cities above 100,000; right side cities 50,000-100,000).*

| $T$ | univ cities | other cities | |
|---|---|---|---|
| Q1 | 14 | 6 | 20 |
| Q2 | 13 | 8 | 21 |
| Q3 | 11 | 9 | 20 |
| Q4 | 6 | 14 | 20 |
| | 44 | 37 | 81 |

p= 0.065

| $T$ | univ cities | other cities | |
|---|---|---|---|
| Q1 | 2 | 25 | 27 |
| Q2 | 2 | 26 | 28 |
| Q3 | 1 | 26 | 27 |
| Q4 | 1 | 27 | 28 |
| | 6 | 104 | 110 |

p= 0.871

| $T$ | univ cities | other cities | |
|---|---|---|---|
| Q1+Q2 | 27 | 14 | 41 |
| Q3+Q4 | 17 | 23 | 40 |
| | 44 | 37 | 81 |

p= 0.035

| $T$ | univ cities | other cities | |
|---|---|---|---|
| Q1+Q2 | 4 | 51 | 55 |
| Q3+Q4 | 2 | 53 | 55 |
| | 6 | 104 | 110 |

p= 0.401

| $U$ | univ cities | other cities | |
|---|---|---|---|
| Q1 | 15 | 5 | 20 |
| Q2 | 13 | 8 | 21 |
| Q3 | 11 | 9 | 20 |
| Q4 | 5 | 15 | 20 |
| | 44 | 37 | 81 |

p= 0.013

| $U$ | univ cities | other cities | |
|---|---|---|---|
| Q1 | 4 | 23 | 27 |
| Q2 | 1 | 27 | 28 |
| Q3 | 1 | 26 | 27 |
| Q4 | 0 | 28 | 28 |
| | 6 | 104 | 110 |

p <0.000

| $U$ | univ cities | other cities | |
|---|---|---|---|
| Q1+Q2 | 28 | 13 | 41 |
| Q3+Q4 | 16 | 24 | 40 |
| | 44 | 37 | 81 |

p= 0.011

| $U$ | univ cities | other cities | |
|---|---|---|---|
| Q1+Q2 | 5 | 50 | 55 |
| Q3+Q4 | 1 | 54 | 55 |
| | 6 | 104 | 110 |

p <0.000

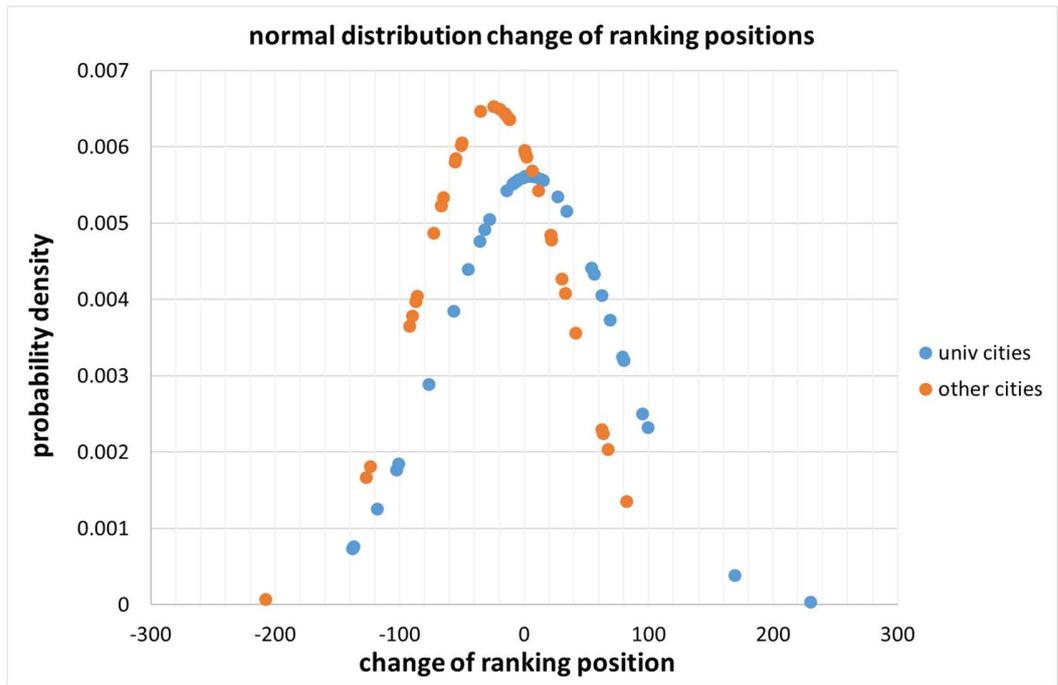

*Fig. S3  Normal distribution of the change in ranking positions 2004-2019 for all German cities above 100,000 inhabitants, with the distinction between university cities (as far as included in the Leiden ranking) and the other cities.*

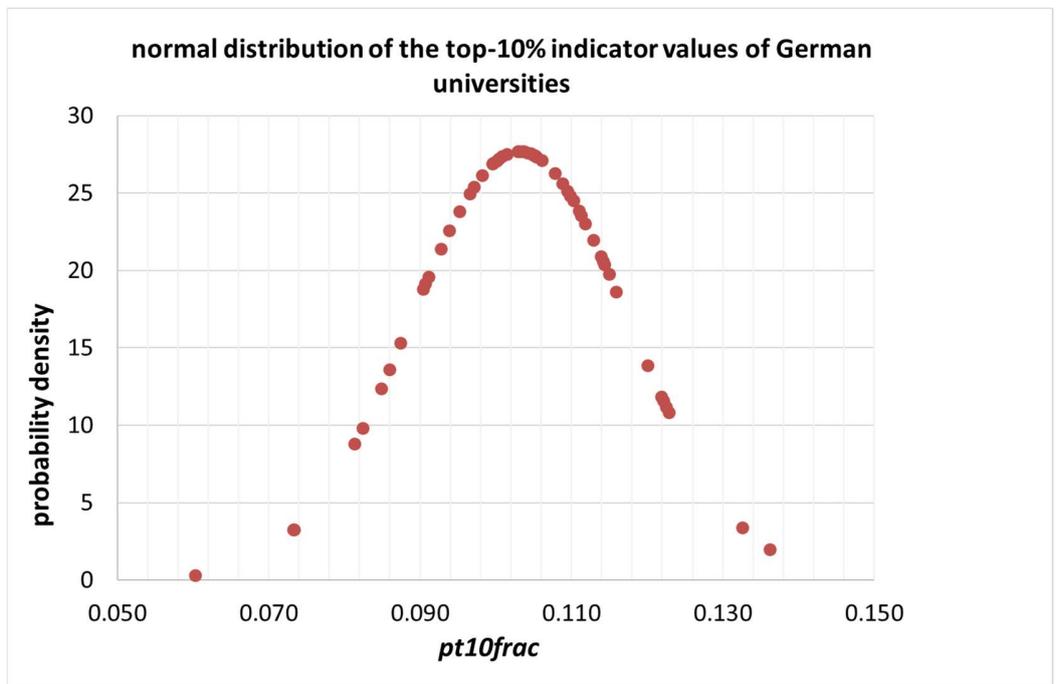

*Fig. S4  Normal distribution of the top-10% indicator (pt10frac) of all German universities as far as included in the Leiden ranking.*

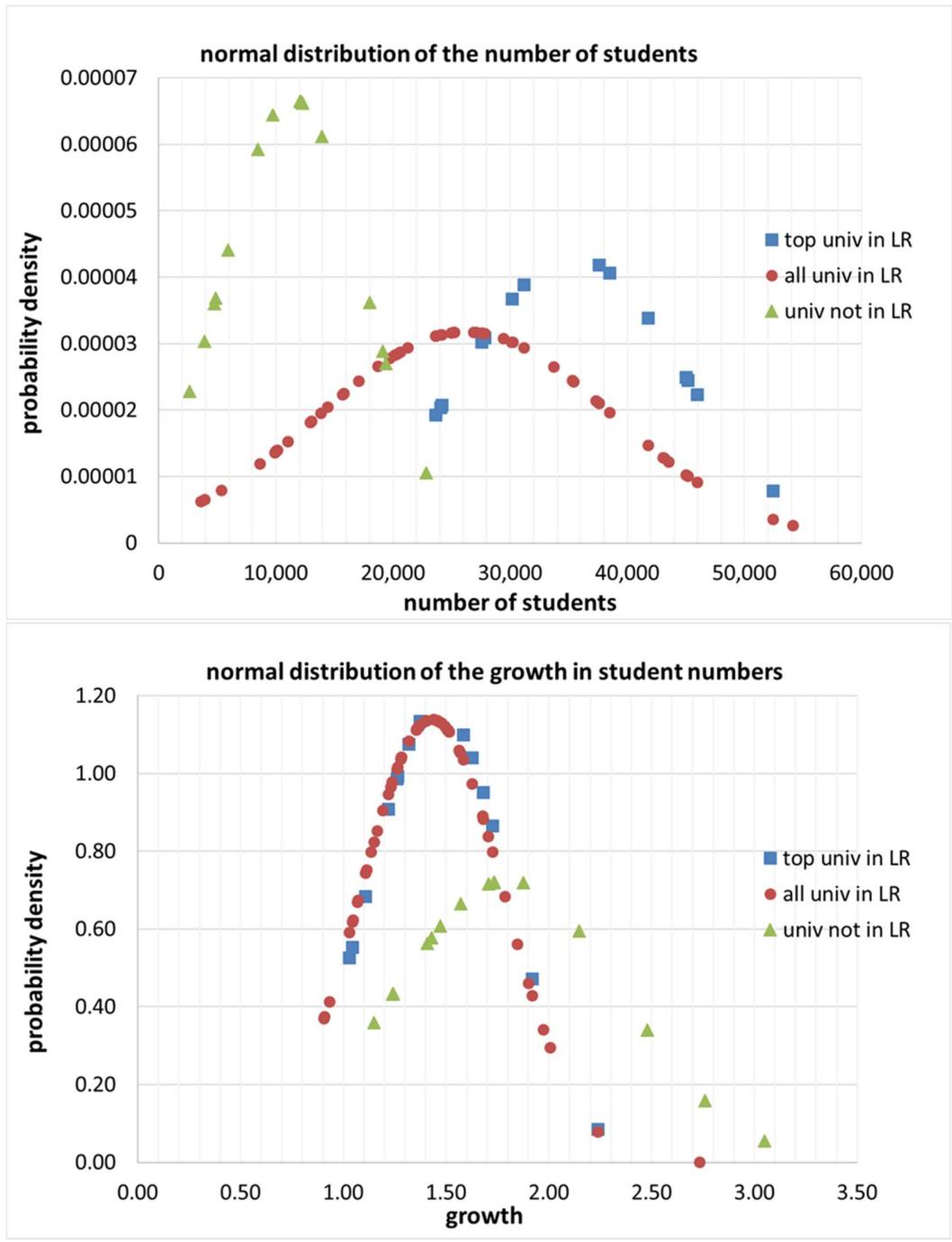

*Fig. S5 Normal distribution of the numbers of students and growth in student numbers for the top-universities in the Leiden Ranking (LR), all LR universities and not-LR universities.*

Table S5. The light blue marked indicators are significantly larger for the first quartile (Q1) or first half (Q1+Q2) as compared with the rest of the $T$ distribution for the 81 cities >100,000.

| $T$(Q1) | Ratio or diff: Q1/(Q2+Q3+Q4) | p | $T$(Q1+Q2) | Ratio or diff: (Q1+Q2)/(Q3+Q4) | p |
|---|---|---|---|---|---|
| $S$ | | | $S$ | | |
| $U$ | 1.09 | 0.001 | $U$ | 1.08 | 0.002 |
| $T$ | 1.22 | 0.000 | $T$ | 1.18 | 0.000 |
| V | | | V | | |
| Pfrac | | | Pfrac | | |
| Pt1frac | | | Pt1frac | | |
| Pt5frac | | | Pt5frac | | |
| Pt10frac | | | Pt10frac | | |
| Pt50frac | | | Pt50frac | | |
| pt1frac | | | pt1frac | | |
| pt5frac | | | pt5frac | | |
| pt10frac | | | pt10frac | | |
| pt50frac | | | pt50frac | | |
| Pfull | | | Pfull | | |
| Pt1full | | | Pt1full | | |
| Pt5full | | | Pt5full | | |
| Pt10full | | | Pt10full | | |
| Pt50full | | | Pt50full | | |
| pt1full | | | pt1full | | |
| pt5full | | | pt5full | | |
| pt10full | | | pt10full | | |
| pt50full | | | pt50full | | |
| Cfrac | | | Cfrac | | |
| Cfull | | | Cfull | | |
| Pcoll | | | Pcoll | | |
| Pintcoll | | | Pintcoll | | |
| Pb | | | Pb | | |
| PbL | | | PbL | | |
| N(s) | | | N(s) | | |

Table S6. Example of the difference in results between method 1 (left table) and method 2 (right table). Cities in the white fields of the left side of the table are in the higher half (Q1+Q2) of the socioeconomic strength ($S$) distribution but they are not in the higher half (Q1+Q2) of the pt10frac distribution (unlike the cities in the blue fields). The other way around, cities in the white fields of the right side of the table are in the higher half (Q1+Q2) of the pt10frac but they are not in the higher half (Q1+Q2) the socioeconomic strength ($S$) distribution (unlike the cities in the blue fields).

| $S$(Q1+Q2) | pt10frac | | pt10frac (Q1+Q2) | pt10frac |
|---|---|---|---|---|
| Göttingen | 0.136 | | Göttingen | 0.136 |
| München | 0.133 | | München | 0.133 |
| Bonn | 0.122 | | Bonn | 0.122 |
| Heidelberg | 0.122 | | Heidelberg | 0.122 |
| Würzburg | 0.122 | | Würzburg | 0.122 |
| Münster | 0.120 | | Münster | 0.120 |
| Mainz | 0.116 | | Mainz | 0.116 |
| Stuttgart | 0.115 | | Stuttgart | 0.115 |
| Frankfurt | 0.114 | | Frankfurt | 0.114 |
| Erlangen | 0.114 | | Erlangen | 0.114 |
| Nürnberg | 0.114 | | Nürnberg | 0.114 |
| Aachen | 0.113 | | Freiburg | 0.114 |
| Karlsruhe | 0.112 | | Aachen | 0.113 |
| Köln | 0.111 | | Karlsruhe | 0.112 |
| Regensburg | 0.110 | | Berlin | 0.111 |
| Darmstadt | 0.109 | | Köln | 0.111 |
| Kaiserslautern | 0.104 | | Regensburg | 0.110 |
| Potsdam | 0.103 | | Duisburg | 0.109 |
| Hamburg | 0.103 | | Essen | 0.109 |
| Düsseldorf | 0.101 | | Darmstadt | 0.109 |
| Jena | 0.101 | | Kassel | 0.106 |
| Paderborn | 0.100 | | Dresden | 0.105 |
| Hannover | 0.100 | | | |
| Braunschweig | 0.095 | | | |
| Ulm | 0.091 | | | |
| Saarbrücken | 0.085 | | | |
| Bremen | 0.082 | | | |